%% file: arxiv_version.tex
\documentclass[a4paper,11pt]{article}

\usepackage{amsmath,amsfonts,amssymb,amsthm,amscd}
\usepackage{dsfont}
\usepackage{graphicx,mathtools}
\usepackage{perpage}
\usepackage{url}
\usepackage{color}
\usepackage[T1]{fontenc}
\usepackage{hyperref}
\usepackage[a4paper,scale={0.72,0.74},marginratio={1:1},footskip=7mm,headsep=10mm]{geometry}
\usepackage[latin9]{inputenc}
\usepackage{mathrsfs}
\usepackage{caption}
\usepackage{subcaption}
\usepackage{float}
\usepackage{enumerate}
\usepackage{natbib}
\usepackage[english]{babel}
\usepackage{hyperref}
\usepackage{setspace}
\usepackage{rotating}

\usepackage{multirow}
\usepackage{subcaption}
\usepackage{diagbox}
\usepackage{booktabs} 
\usepackage{threeparttable}
\usepackage[ruled]{algorithm2e} 

\makeatletter
\def\@captype{figure}
\makeatother

\numberwithin{equation}{section}

\newtheorem{theorem}{Theorem}[section]

\newtheorem{conjecture}[theorem]{Conjecture}

\DeclareMathOperator*{\argmax}{arg\,max}
\DeclareMathOperator*{\argmin}{arg\,min}

\begin{document}


\title{Equilibrium stability as a driver of cooperation among Q-learners\\
}

\author{
\renewcommand{\thefootnote}{\arabic{footnote}}
Janusz M.\ Meylahn
\footnotemark[1]
\\
\renewcommand{\thefootnote}{\arabic{footnote}}
Maximilian Sch\"{a}fer
\footnotemark[2]
\\
}

\footnotetext[1]{
University of Twente, Department of Applied Mathematics, Drienerloolaan 5, 7522 NB Enschede, The Netherlands. \href{mailto:j.m.meylahn@utwente.nl}{j.m.meylahn@utwente.nl}\vspace{0cm}
}

\footnotetext[2]{
Institut Mines-T\'{e}l\'{e}com Business School, Department of Data Analytics, Economics, and Finance. 9 Rue Charles Fourier, 91000 Evry, France. \href{mailto:maximilian.schafer@imt-bs.eu}{maximilian.schafer@imt-bs.eu}
}

\date{\today}

\maketitle


\begin{abstract}
Algorithmic collusion among pricing algorithms has raised concerns about sustained supra-competitive prices and their implications for social welfare. Existing work has largely focused on the probability that reinforcement-learning algorithms converge to cooperative strategies, typically under the assumption that exploration vanishes over time. Motivated by the observation that algorithms deployed in practice are likely to continue exploring in order to remain adaptive to changing environments, we study learning dynamics under constant exploration. In this setting, the relevant question is no longer whether an algorithm converges to a particular strategy profile, but rather what fraction of time the algorithms spend playing cooperative strategies. Even in the benchmark case of the repeated Prisoner's Dilemma with one-period memory, this yields high-dimensional stochastic learning dynamics, for which a complete analytic treatment is intractable. We show that cooperative strategies can be dominant in this time-averaged sense and derive a boundary predicting when such dominance arises, based on the expected dynamics of the Q-learning process. Extensive simulations show that this boundary is a strong predictor for non-defection-dominated behaviour under epsilon-greedy Q-learning.



\medskip\noindent
{\it Artificial Intelligence, Pricing-Algorithms, Collusion, Reinforcement Learning, Q-Learning} 


\end{abstract}

\newpage
\section{Introduction}
\label{sec:intro}

A large amount of recent work has highlighted the potential danger of algorithmic collusion among pricing algorithms \citep{assad2024algorithmic, brown2023competition}. Such collusion would lead to supra-competitive prices and is most likely to be legal under current antitrust law \citep{mackay2022dynamic, harrington2018developing}. Early work already showed that reinforcement-learning algorithms can coordinate on supra-competitive outcomes in oligopoly settings \citep{waltman2008q}, while more recent studies have shown that Q-learning algorithms may sustain elevated prices in combination with reward-punishment schemes consistent with collusive behaviour \citep{calvano2020artificial, klein2021autonomous}; alongside proposals for auditing algorithms for collusion \citep{hartline2024regulation}. Whether algorithmic collusion poses a realistic threat to social welfare remains debated, with multiple studies reporting mixed evidence and questioning the robustness and interpretation of such cooperative outcomes \citep{ asker2022artificial, lambin2024less}.

Much of the reinforcement-learning literature studies algorithms with a constant learning rate and allows for persistent stochasticity in the environment. This modelling choice is standard and well motivated: a constant learning rate ensures that the algorithm remains responsive to changes or non-stationarities in payoffs, while environmental noise guarantees continued exploration of the state space \citep{kushner1981asymptotic, kushner1981averaging, beck2012error}. In this setting, stochasticity is exogenous to the algorithm and serves to support adaptability.

By contrast, in the algorithmic collusion literature, stochasticity typically originates from the agents' own exploration. Many studies assume that exploration rates decline over time, so that learning eventually becomes effectively deterministic \citep{calvano2020artificial, calvano2021algorithmic, asker2022artificial}. While this facilitates convergence to stable pricing policies, it also reduces the ability of algorithms to adapt to changes in the environment or to the behaviour of other learning agents. This limitation is particularly salient in multi-agent settings, which are inherently non-stationary \citep{denBoer2022artificial, den2024mathematical, lambin2024less}. Moreover, real-world demand conditions are unlikely to be stable over long horizons, suggesting that pricing algorithms deployed in practice may need to continue exploring in order to remain effective.

In this paper, we study cooperation among Q-learners with a memory of one period in the prisoner's dilemma when both the exploration and learning rates are kept constant. Our focus on memory-one strategies builds on recent work characterising equilibrium structure and learning dynamics in repeated prisoner's dilemma games \citep{usui2021symmetric, meylahn2022limiting}. With constant exploration and learning, the resulting learning dynamics no longer converge to a particular strategy profile and stay there indefinitely. Instead, the system transitions persistently between strategy profiles. This observation motivates a shift in perspective: rather than asking whether learning converges to a cooperative outcome, we measure cooperation by the fraction of time the learning dynamics spend in cooperative versus non-cooperative strategy profiles. 

We provide a theoretical characterisation of these learning dynamics and propose a heuristic boundary predicting when cooperation dominates in this time-averaged sense. The boundary is derived from the relative stability of equilibrium strategy profiles in the presence of persistent fluctuations. Related work has shown that stochastic fluctuations can play a central role in sustaining or disrupting cooperation in learning and evolutionary dynamics, including in repeated prisoner's dilemma settings \citep{barfuss2023intrinsic,dolgopolov2024reinforcement,xu2024mechanism}. This stability-based mechanism uncovers a novel route to cooperation among reinforcement-learning algorithms--one that does not rely on vanishing exploration or absorption into a cooperative equilibrium. Extensive simulations show that the proposed boundary predicts the empirical transition between cooperative and non-cooperative regimes with high accuracy

This perspective complements existing work in the algorithmic collusion literature, where collusion is typically assessed by the probability of converging to supra-competitive pricing once learning has effectively ceased. We introduce a new perspective by focusing on the long-run fraction of time that algorithms price supra-competitively as the measure of collusion. From a regulatory standpoint, this distinction is important because the extent of harm to social welfare is a function not only of the price but also of the time that the price is used. 

Our results show that equilibrium stability can be used to predict the fraction of time spent pricing supra-competitively. Interpreted economically, this stability reflects the robustness of algorithmic cartels to ongoing fluctuations induced by learning and exploration. Social welfare losses from algorithmic collusion thus depend not only on the existence of collusive strategies, but on their resilience to persistent noise in the learning process.

We approach the problem by considering the joint dynamics of the state and Q-values as a Markov process. Here the state space is discrete while the Q-value space is continuous. Analysing the true dynamics of this process is intractable, but by projecting from the Q-value to the strategy profile space, we obtain a hidden Markov model with a finite state space. Our observable of interest is the distribution of occupation times over strategy profiles. We prove that there is a positive probability of observing a transition from any strategy profile to another strategy profile, where each of the strategies is a best response to some strategy. Strategy profiles with this property thus form part of a unique recurrent set of the dynamics. Based on this proof, we conjecture  that the distribution of occupation times converges to a unique stationary distribution in the long time limit.

By considering an idealised learning process where we only take into account the expected learning dynamics and linking this to properties of the unique stationarity distribution, we derive an analytic condition on the parameters for predicting when the stationary distribution places more mass on cooperative strategy profiles than on competitive strategy profiles. Concretely, this is based on comparing the optimal Q-value differences at different stationary points of the expected learning dynamics. This comparison captures the idea that the stability of a strategy profile is predominantly determined by the probability of observing a reversal in the order of the Q-values. The probability of such a reversal is higher when the optimal Q-value difference is smaller. 

Even though the stationary distribution of the occupation time is reached in the long time limit, the time it takes to be stationary depends on the choice of parameters. Most important in this regard are the learning and exploration rates. For our simulations, we thus fix a time horizon and identify the regions of the parameter space for which stationarity is reached within this time horizon. This gives us the region in which we can expect our theoretical framework to apply. We proceed by sweeping the parameter space and recording the average empirical occupation time in the stationary distribution for the strategy profiles that occur most commonly in the simulations. 

We compare the theoretical boundary to the simulation outcomes by overlaying it on heatmaps of average empirical occupation times. This visual comparison shows that the boundary closely aligns with the transition between cooperative and non-cooperative strategy profiles. To provide a more systematic evaluation, we interpret the boundary as a decision rule to predict the dominance of cooperative outcomes and assess its performance using standard statistical metrics for classification. Despite being based on a heuristic that abstracts from several intricacies of the learning dynamics, the boundary performs remarkably well, achieving macro F1 scores ranging from 0.85 to 0.95.

\section{Related Literature}
\label{sec:relatedlit}

Here, we provide a targeted review of the literature, highlighting its relevance to debates on algorithmic collusion in economics and its connection to, and extension of, prior work on algorithmic interactions in the prisoner's dilemma and on stochastic effects in reinforcement-learning-based cooperation.

\paragraph{Economic Literature on Algorithmic Collusion.}
A series of studies has nuanced the interpretation of simulation-based evidence of algorithmic collusion \citep{asker2024impact, abada2023artificial, epivent2024algorithmic, abada2024collusion}. More specifically, \citet{abada2023artificial} and \citet{lambin2024less} identify vanishing exploration schedules as a key driver of cooperative patterns in algorithmic pricing environments, arguing that observed reward-punishment schemes may be artefacts of the learning process. This perspective calls into question the economic interpretation of reward-punishment schemes emphasised by \citet{calvano2020artificial} and \citet{klein2021autonomous}. Motivated by this critique, we study a setting with persistent exploration, in which learning does not converge to a single policy and cooperative behaviour must instead be assessed through time-averaged dominance rather than convergence. Within this perspective, we show that cooperative strategy profiles can dominate long-run behaviour and derive a predictive boundary based on the relative stability of equilibrium strategy profiles under ongoing fluctuations. In particular, our analysis compares the stability of a forgiving reward-punishment strategy (win-stay, lose-shift) with that of all-defection, yielding a mechanism for sustained cooperation that does not rely on exploration vanishing to zero.
\paragraph{Algorithmic interactions in memory-one prisoner's dilemma games.} Recent work has focused on analysing the dynamics of stateless Q-learners in the prisoner's dilemma under a variety of self-play (in the sense of coupling between Q-tables) assumptions \citep{banchio2022artificial, banchio2023adaptive}. The case where the players make use of a memory of one period as a state space has received significant attention recently. Conditions for the existence of equilibria have been derived \citep{usui2021symmetric, meylahn2022limiting}. Convergence and a basin of attraction analysis of a batched version of Q-learning have been shown for the prisoner's dilemma \citep{meylahn2025quantifying} and a three-action variant thereof \citep{meylahn2023does}. The metagame of hyperparameter selection has been analysed \citep{carissimo2025algorithmic}, and conditions for the transition from defection to cooperation have been derived under a self-play assumption \citep{bertrand2025self}. Previous work has also focused on the relationship between the speed of learning and exploration rate decrease \citep{denBoer2022artificial}.
\paragraph{Fluctuations.}
The papers most closely related to our work are those that examine the interaction between fluctuations and cooperation. Two recent studies employ the framework of stochastic stability to analyse mechanisms of algorithmic cooperation in the prisoner's dilemma without memory \citep{dolgopolov2024reinforcement, xu2024mechanism}. Closely related is the finding that intrinsic fluctuations can facilitate cooperation in batched versions of Q-learning \citep{barfuss2023intrinsic}. An alternative modelling approach treats actions as choices among memory-one strategies played for fixed periods, with reinforcement learning used to adapt strategy selection. When the strategy set includes Tit-for-Tat, Always-Defect, and Always-Cooperate, stochasticity can induce cycles between cooperation and defection \citep{galla2011cycles}. Similar dynamics arise when the Win-Stay, Lose-Shift strategy is included \citep{bladon2010evolutionary}. Our heuristic boundary builds on these fluctuation-based insights by identifying the Q-value differences that are critical for assessing the relative stability of competing equilibria.

\section{Model}
\label{sec:setting}
In this section, we introduce the environment and algorithm we will consider. Together they determine a Markovian process. To facilitate our analysis in Section~\ref{sec:theory}, we also introduce a related idealised process.

The environment we consider is the iterated prisoner's dilemma, where two players, labelled $i\in\{1, -1\}$, employ a state space which consists of all one-period histories and can thus condition their action-selection probabilities on the actions taken in the previous round. This means that we have $\mathcal{S}=\{(D, D), (D, C), (C, D), (C, C)\}$ and $\mathcal{A}_{i}=\mathcal{A}=\{D, C\}$, where $D$ denotes the ``defect'' action and $C$ denotes the ``cooperate'' action. We denote the state at time $t\in\{0, 1, \ldots\}$ by $s_{t}\in\mathcal{S}$ and the actions taken at time $t$ by $\boldsymbol{a}_{t}=(a^{1}_{t}, a^{-1}_{t})$, with $a_{t}^{i}$ the action of player $i$. Note that $s_{t+1}=\boldsymbol{a}_{t}$, which thus defines our transition dynamics. The rewards $\boldsymbol{r}(\boldsymbol{a})=(r^{1}(\boldsymbol{a}), r^{-1}(\boldsymbol{a}))$ are independent of the state and are defined as 
\begin{align}
r^{1}(\boldsymbol{a}) = 
    \begin{cases}
        P \quad \text{when } \boldsymbol{a} = (D, D)\\
        T \quad \text{when } \boldsymbol{a} = (D, C)\\
        S \quad \text{when } \boldsymbol{a} = (C, D)\\
        R \quad \text{when } \boldsymbol{a} = (C, C)\\
    \end{cases},
    \quad\text{and}\quad
r^{-1}(\boldsymbol{a}) = 
    \begin{cases}
        P \quad \text{when } \boldsymbol{a} = (D, D)\\
        S \quad \text{when } \boldsymbol{a} = (D, C)\\
        T \quad \text{when } \boldsymbol{a} = (C, D)\\
        R \quad \text{when } \boldsymbol{a} = (C, C)\\
    \end{cases},
\end{align}
with $T>R>P>S$. The players want to maximise the expected sum of discounted rewards, i.e., 
\begin{equation}
    \mathbb{E}\Bigg[\sum_{t=1}^{\infty}\delta^{t}r_{t}^{i}\Bigg],
\end{equation}
where $\delta\in(0, 1)$. We are interested in studying the dynamics of the system 
\begin{equation*}
    Q_{t+1}^{i}(s, a) = 
    \begin{cases}
        (1-\alpha)Q_{t}^{i}(s, a) + \alpha[(1-\delta)r_{t}(\boldsymbol{a}) &\text{for } (s, a) = (s_{t}, a_{t}^{i}) \text{ and } \boldsymbol{a}=\boldsymbol{a}_{t}\\
         \hspace{2cm} +\; \delta \max_{a'}Q_{t}^{i}(s_{t+1}, a')] \\
        Q_{t}^{i}(s, a) \quad &\text{else}   
    \end{cases},
\end{equation*}
where we initialise the Q-values $Q_{0}^{i}$ as $U[S, T]$ (uniformly between $S$ and $T$), $\alpha\in(0, 1)$ is the learning rate, and we normalise the payoffs by $1-\delta$ to ensure that the Q-values remain in $\mathcal{Q} =[S, T]$ for aesthetic purposes and without loss of generality. At each point in time, player $i$ selects an action according to the $\epsilon$-greedy mechanism, with $\epsilon\in(0, 1)$
\begin{equation}
    a_{t}^{i} = 
    \begin{cases}
        \argmax_{a\in\mathcal{A}}Q_{t}^{i}(s_{t}, a) \quad &\text{with probability } 1-\epsilon/2\\
        \argmin_{a\in\mathcal{A}}Q_{t}^{i}(s_{t}, a) \quad &\text{with probability } \epsilon/2
    \end{cases}.
\end{equation}
This means that a player plays their greedy action with probability $1-\epsilon$ and takes an action uniformly at random with probability $\epsilon$. Our setting is thus parameterised by $\theta = (T, R, P, S, \delta, \epsilon, \alpha) \in \Theta$, where $\Theta$ denotes the space of allowed values for the parameters. We can split $\theta$ into the environmental parameters $\theta^{e}=(T, R, P, S)$ and the algorithmic hyperparameters $\theta^{a}=(\delta, \epsilon, \alpha)$, such that $\theta=(\theta^{e}, \theta^{a})$. The underlying Q-value dynamics together with the action-selection mechanism lead to trajectories of state-action pairs
\begin{equation}
    \{s_{0}, \boldsymbol{a}_{0}, s_{1}, \boldsymbol{a}_{1}, \ldots \},
\end{equation}
where $s_{0}$ is taken uniformly from $\mathcal{S}$. Since we have $s_{t+1} =\boldsymbol{a}_{t}$, it is sufficient to consider
\begin{equation}
  \{s_{0}, s_{1}, s_{2}, \ldots \}.
\end{equation}
The state of the system as a whole consists of 
\begin{equation}
    (s_{t}, \boldsymbol{Q}_{t}) \in \mathcal{S}\times \mathcal{Q}\times\mathcal{Q},
\end{equation}
where $\boldsymbol{Q}_{t} = (Q_{t}^{1}, Q_{t}^{-1})$.  The Q-table of a player can be projected onto the space of $\epsilon$-greedy strategies $\Pi_{\epsilon}$ with a fixed $\epsilon$ using
\begin{align*}
    \pi(Q^{i}_{t}) = \Bigg(&\argmax_{a\in\mathcal{A}}Q_{t}^{i}((D, D), a), \argmax_{a\in\mathcal{A}}Q_{t}^{i}((D, C), a), \\
    &\argmax_{a\in\mathcal{A}}Q_{t}^{i}((C, D), a), \argmax_{a\in\mathcal{A}}Q_{t}^{i}((C, C), a)\Bigg).
\end{align*}
For example, if player $i$ has $Q_{t}^{i}(s, D)>Q_{t}^{i}(s, C)$ for all $s\in\mathcal{S}$, then they are playing the $\epsilon$-greedy All-Defect (AD) strategy with $\pi(Q^{i}) = (D, D, D, D)$. In our setting, there are $|\Pi_{\epsilon}|=16$ such strategies and therefore 256 strategy profiles $\boldsymbol{\pi}\in\Pi_{\epsilon}\times \Pi_{\epsilon}$.

We will predominantly be studying the dynamics of 
\begin{equation}
    (s_{t}, \boldsymbol{\pi}(\boldsymbol{Q}_{t}))\in \mathcal{S}\times \Pi_{\epsilon}\times \Pi_{\epsilon}.
\end{equation}
Note that the dynamics of $(s_{t}, \boldsymbol{Q}_{t})$ are Markovian, but that the dynamics of $(s_{t}, \boldsymbol{\pi}(\boldsymbol{Q}_{t}))$ are not, since the probability of observing a particular transition between $\epsilon$-greedy strategy profiles depends on the exact Q-values. We are thus studying a hidden Markov model. The advantage of studying $(s_{t}, \boldsymbol{\pi}(\boldsymbol{Q}_{t}))$ is that these dynamics occur in a discrete space. 

\paragraph{Idealised Process.} To derive our boundary predicting cooperation, we will analyse points of the dynamics at which the expected dynamics are stationary. To identify these, we make use of an idealised process, which we describe next. At every point in time, we can consider a single player $i$ to be facing a stationary environment defined by the combination of the prisoner's dilemma and the opponent's $\epsilon$-greedy strategy $\pi_{t}^{-i}$. This stationary environment defines an optimal best response $\text{BR}(\pi_{t}^{-i})$ which is unique for almost all\footnote{The best response is not unique for a set of parameters with Lebesgue measure zero.} $T, R, P, S, \delta$, and $\epsilon$. In an idealised learning process, each agent would have perfect information of this stationary environment and could thus update their strategy to or towards the best response at each learning step.  

The best-response function $\text{BR}(\cdot)$ induces a functional relation on $\Pi_{\epsilon}$, called the Individual Best-Response (IBR) graph, in which Nash equilibria (NE) appear as self-loops or two-cycles \citep{meylahn2025quantifying}. In previous work, it was shown that in the prisoner's dilemma with a memory of one period there are three possible NE \citep{usui2021symmetric,meylahn2022limiting}, namely, both players playing 
\begin{equation}
    \pi^{\text{AD}} = (D, D, D, D), \quad \pi^{\text{GT}}=(D, D, D, C) \;\text{ or }\; \pi^{\text{WSLS}} = (C, D, D, C).
\end{equation}
Each NE is associated with optimal Q-values, which we name $Q^{\text{AD}}, Q^{\text{GT}}$ and $Q^{\text{WSLS}}$, respectively. These are defined as the solution to Bellman's optimality equation when both players are playing the strategy associated with the NE, i.e., the solution to 
\begin{equation}
    Q^{\text{NE}}(s, a;\theta) = \mathbb{E}_{\text{NE}}\Big[(1-\delta)r^{i}(\boldsymbol{a}) + \delta \max_{a\in\mathcal{A}}Q^{\text{NE}}(s', a;\theta)\Big],
\end{equation}
where the expectation is with respect to the dynamics induced by both players playing $\text{NE}$ and $s'$ is the next state. Here we make the dependence on the parameters $\theta$ explicit. For both the idealised and true learning process, the expected dynamics of $\boldsymbol{Q_{t}}$ are stationary at these values, that is, 
\begin{equation}
    \mathbb{E}[\boldsymbol{Q}_{t+1}|\boldsymbol{Q}_{t} = (Q^{\text{NE}}, Q^{\text{NE}})] =  (Q^{\text{NE}}, Q^{\text{NE}}),
\end{equation} 
for $\text{NE}\in\{\text{AD}, \text{GT}, \text{WSLS}\}$.
 
\section{Theoretical Analysis}
\label{sec:theory}

In this section, we give our theoretical analysis of the dynamics of the model described in Section~\ref{sec:setting}. The Markovian dynamics of $(s_{t}, \boldsymbol{Q}_{t})$ are not necessarily ergodic, since it could be that there are regions of the Q-value space $\mathcal{Q}$ that cannot be reached once they have been left\footnote{Adding, for example, Gaussian noise to the rewards would likely make the dynamics ergodic.} \citep{meyn2012markov}.

We are interested in studying the fraction of time spent in each of the strategy profiles and the fraction of time spent in each of the four states. To this end, we introduce the empirical occupation time of strategy profiles for a trajectory up to time horizon $T_{h}$
\begin{equation}
    \tau_{T_{h}} = \{\boldsymbol{\pi}_{0}, \boldsymbol{\pi}_{1}, \ldots \boldsymbol{\pi}_{T_{h}}\}
\end{equation}
as
\begin{equation}
\label{eq:occupationprofiles}
    O(\tau_{T_{h}}, \boldsymbol{\pi}) = \frac{1}{T_{h}}\sum_{t=1}^{T_{h}}\mathds{1}_{\{\boldsymbol{\pi_{t}=\boldsymbol{\pi}}\}}.
\end{equation}
$O(\tau_{T_{h}}, \boldsymbol{\pi})$ captures the fraction of time that was spent in $\boldsymbol{\pi}$ during trajectory $\tau_{T_{h}}$. Similarly, and with a slight abuse of notation,\footnote{The second argument of the function makes clear which occupation time is meant.} we define the empirical occupation time of the states to be 
\begin{equation}
\label{eq:occupationstates}
    O(\sigma_{T_{h}}, s) = \frac{1}{T_{h}}\sum_{t=1}^{T_{h}}\mathds{1}_{\{s_{t}=s\}},
\end{equation}
with $\sigma_{T_{h}}=\{s_{0}, s_{1}, \ldots, s_{T_{h}}\}$. 

For Conjecture~\ref{thm:stationarydist} we argue that the occupation time of strategy profiles converges to a unique stationary distribution when $\alpha$ is small enough. If this is the case, the distribution over states will also be stationary and unique. 

\begin{theorem}[Unique Recurrent Set]
\label{thm:recurrentset}
    Fix $T, R, P, S, \delta$, and $\epsilon$, then there exists $\alpha^{c}$ such that for $\alpha<\alpha^{c}$ the set of strategy profiles
    \begin{equation}
        \{(\tilde{\pi}^{1}, \tilde{\pi}^{-1}):\tilde{\pi}^{i} = BR(\pi) \text{ for some } \pi\in\Pi_{\epsilon}\} 
    \end{equation}
    form part of a unique recurrent set.
\end{theorem}
The proof of Theorem \ref{thm:recurrentset} is given in Appendix \ref{app:proof}, but we provide a brief sketch of the proof here. Starting at any $(s_{t}, \boldsymbol{Q}_{t})$, we observe any finite sequence of play of length $K$ with positive probability. In particular, we may observe a sequence that looks as if it were generated by the players playing strategy profile $(\pi^{1}, \pi^{-1})$ for any $\pi^{1}, \pi^{-1}\in\Pi$. From the perspective of player $i$, it thus seems as if they are facing an opponent playing $\pi^{-i}$ for those $K$ periods. Using \cite{beck2012error}, we can establish that the Q-values that player $i$ learns during this sequence of play are sufficiently close to the Q-values corresponding to $BR(\pi^{-i})$ that the player actually learns the best response. This occurs with positive probability when $\alpha$ is less than some critical value, which we calculate explicitly. Since this occurs with positive probability in a finite number of steps for any initial condition, we can split the time horizon into non-overlapping blocks of length $K$ to conclude that we will return to all strategy profiles consisting of strategies that are a best response to some other strategy infinitely often. Note that there may be more strategy profiles in the unique recurrent set than the ones identified here.

\begin{conjecture}[Unique Stationary Distribution]
\label{thm:stationarydist}
    Fix $T, R, P, S, \delta$, and $\epsilon$, then there exists $\alpha^{c}$ such that for $\alpha<\alpha^{c}$
    \begin{equation}
        \lim_{T_{h}\rightarrow\infty}O(\tau_{T_{h}}, \boldsymbol{\pi}) = \rho(\theta, \boldsymbol{\pi})
    \end{equation}
    for all $\boldsymbol{\pi}$ and $\boldsymbol{\pi}_{0}$ and some $\rho(\theta, \boldsymbol{\pi})$.
\end{conjecture}
From Theorem \ref{thm:recurrentset} we know that the Q-values will return to the neighbourhood of a set of Q-values corresponding to strategy profiles of a particular kind infinitely often when $\alpha<\alpha_{c}$, but this does not immediately imply that a stationary distribution for the Markov process exists and is unique. A potential approach for showing that this is the case would be to use the Krylov-Bogolioubov Theorem \cite{Kryloff1937} to show that there exists \textit{at least one} invariant probability measure of the process and then to use \cite[Corollary 2.7]{hairer2010convergence} to show that there exists \textit{at most} one invariant measure. To apply these two results, the process must (1) take place in a Polish space, (2) there must be at least one accessible point, (3) there must exist an initial condition $x_{0}=(s_{0}, \boldsymbol{Q}_{0})$ so that the sequence $\mathcal{P}^{n}(x_{0}, \cdot)$ is tight, and (4) the operator $\mathcal{P}$ of the process must be a strong Feller Markov operator. Now (1) is satisfied, (2) also holds as shown in the proof of Theorem \ref{thm:recurrentset} and (3) holds as the space is compact. It remains to be shown that $\mathcal{P}$ is a strong Feller operator. Verifying this condition is more involved, and it is unclear whether it holds.\footnote{Note that the conjecture might be true even if the fourth condition doesn't hold, as then other proof strategies are available.} However, our simulations suggest that Conjecture~\ref{thm:stationarydist} holds.

\begin{figure}[h!]
    \centering
    \includegraphics[width=0.45\linewidth]{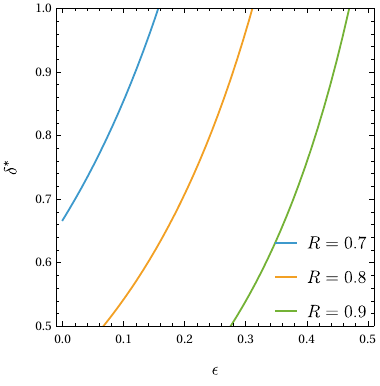}
    \includegraphics[width=0.45\linewidth]{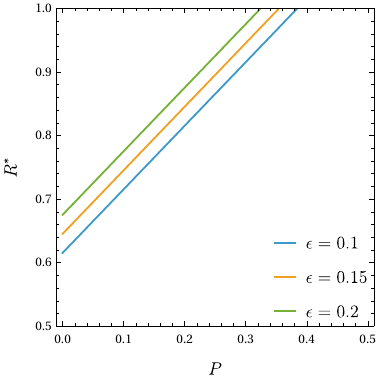}
    \caption{Critical condition for the stability of WSLS being greater than that of AD. On the left, we plot the critical condition on $\delta$ as a function of $\epsilon$ for different values of $R$ (with $T=1$, $S=0$ and $P=0.1$), and on the right we plot the critical condition on $R$ as a function of $P$ for different values of $\epsilon$ (with $T=1$, $S=0$ and $\delta=0.8$). }
    \label{fig:criticalcond}
\end{figure}

The question we are interested in is when the fraction of time spent in cooperative strategy profiles dominates. Of the three NE, only WSLS leads to stable cooperation. So if $\rho(\theta, \boldsymbol{\pi})$ only places positive mass on NE\footnote{This premise is not satisfied in our case, but the argument serves as a basis for deriving our heuristic boundary.}, we would want to know when
\begin{equation}
    \rho(\theta, \pi^{\text{WSLS}}) > 0.5.
\end{equation}
We will proceed heuristically to find a critical condition $\theta^{c}$ on the parameters that predicts when cooperative strategy profiles dominate the stationary distribution. To this end, note that the fraction of time spent in the three equilibria depends on their stability. The stability of an equilibrium depends on at least two components \citep{Freidlin2012}:
\begin{enumerate}
    \item the size of the fluctuations in the equilibrium,
    \item and the height of the barrier that the fluctuations must overcome to exit the equilibrium. 
\end{enumerate}
For our boundary, we will focus only on the second component. To motivate this, note that the exploration rate, which is the source of fluctuations, is the same in all equilibria.\footnote{This does not mean that the fluctuations in Q-value estimates are the same as these also depend on the rewards and continuation payoffs.} It is thus reasonable to assume that the difference in stability between equilibria is predominantly determined by the second component. To exit an equilibrium, the Q-value orderings must be reversed in one of the four states. Intuitively, this is most likely to occur in the most vulnerable state, by which we mean the state with the smallest optimal Q-value difference, and the height of the barrier is characterised by the optimal Q-value differences in that state. For each equilibrium, we thus define the Q-value differences in each state as
\begin{equation}
    \Delta Q^{\text{NE}}(s;\theta) := Q^{\text{NE}}(s, \argmax_{a\in\mathcal{A}} Q^{\text{NE}}(s, a;\theta);\theta) - Q^{\text{NE}}(s, \argmin_{a\in\mathcal{A}} Q^{\text{NE}}(s, a;\theta);\theta).
\end{equation}
To determine when cooperation will be dominant, we want to know when the WSLS equilibrium is more stable than the AD equilibrium\footnote{Note that AD is always a Nash equilibrium, while WSLS is only a Nash equilibrium in some parts of the parameter space \citep{meylahn2025quantifying}.}. Our critical condition is thus defined by solving 
\begin{equation}
    \min_{s\in\mathcal{S}}\Delta Q^{\text{WSLS}}(s;\theta) = \min_{s\in\mathcal{S}}\Delta Q^{\text{AD}}(s;\theta)
\end{equation}
for $\theta$. Solving this, we find the critical condition for the discount factor
\begin{equation}
\label{eq:boundary}
    \delta^{c} := \frac{2(T+P - (R+S))}{(1-\epsilon)[2(R-P)+\epsilon(P+S - (R+T))]},
\end{equation}
such that WSLS is more stable than AD when $\delta>\delta^{c}$. The details of this derivation are given in Appendix~\ref{app:criticalboundary}. In Figure~\ref{fig:criticalcond}, we plot the critical condition in two slices of the parameter space. The boundary behaves sensibly in the following sense. $\delta^{c}$ increases as $\epsilon$ increases and as $R$ decreases, meaning that the required discount factor for cooperation dominating increases as the exploration rate increases and as the value of cooperation decreases. On the other hand, $R^{c}$ increases as $P$ increases and as $\epsilon$ increases. Note that the critical boundary derived here does not depend on the learning rate $\alpha$. However, $\alpha$ does play a role in determining when our boundary applies on the timescale of our simulations (see Section~\ref{sec:accessiblepara}). 

\section{Simulations}
\label{sec:simulations}

\subsection{Simulation Details}
\label{sec:simsetup}

In this subsection, we describe the payoff normalisation used in the simulations, specify which strategy profiles are tracked explicitly, and explain how interaction dynamics outside this focal set of tracked strategies are incorporated.

\subsubsection*{Payoff Normalisation}

Throughout our simulations, we normalise payoffs by setting $T=1$ and $S=0$. This normalisation imposes natural bounds on the remaining payoff parameters $R$ and $P$, allowing us to simulate the full range of incentive structures consistent with the prisoner's dilemma on a bounded parameter grid, which substantially improves computational tractability.

\subsubsection*{Focal Strategy Profiles}

Rather than tracking occupation times for all $256$ possible memory-one strategy profiles, we focus on a small set of theoretically and empirically salient profiles. Specifically, we consider the three Nash equilibrium strategy profiles emerging from our theoretical analysis, together with two additional profiles that appear frequently in preliminary simulations: All-Cooperate ($AC$) and Anti-Grim Trigger ($AGT$), in which both players cooperate following mutual defection and defect otherwise. This set defines the focal strategy class $\hat{\Pi}_{\epsilon}$.

The empirical relevance of $AC$ and $AGT$ can be summarised by the following stylised facts:

\begin{enumerate}
    \item \textbf{Best-response prevalence.} Both $AC$ and $AGT$ arise as best responses to other strategies over nontrivial regions of the parameter space. In particular, $AC$ can be a best response to $GT$, $(D,C,D,D)$, $(D,C,D,C)$, $(D,C,C,C)$, and $(C,C,D,C)$, while $AGT$ can be a best response to $(D,C,C,C)$ (see Table~IV in \cite{meylahn2025quantifying}). By Theorem~\ref{thm:recurrentset}, this implies a positive probability of reaching mutual play of $AC$ or $AGT$.

    \item \textbf{Proximity to the stag-hunt region.} Both strategies occur predominantly when $R$ is close to $T$. In this region, the game approaches a stag hunt, in which both $AC$ and $AGT$ are Nash equilibria \cite{meylahn2025quantifying}. When $R \approx T$, learning dynamics may therefore blur distinctions between underlying game classes, since Q-values encode only approximations of expected continuation payoffs.

    \item \textbf{Game-class boundaries.} The strategy $AGT$ rarely appears when $P$ is close to zero, where the game approaches a snowdrift game in which $AGT$ is not an equilibrium. By contrast, when $P$ is close to zero and $R$ is close to one, the game approaches a harmony game, in which $AC$ is always an equilibrium, explaining its prevalence in that region of the parameter space.
\end{enumerate}

\subsubsection*{Tracking Interaction Dynamics Outside the Focal Strategy Profiles}

Our focus on a restricted set of strategy profiles is guided by two considerations: alignment with the theoretical analysis, which compares stability properties of specific profiles, and the desire to capture a substantial share of empirically observed behaviour while maintaining computational and representational tractability.

At the same time, the total fraction of time spent in the five focal profiles varies across the parameter space.\footnote{Table~\ref{tab:fstrat_sum} shows part of this variation and confirms that the five focal strategies dominate in the stationary distribution throughout our simulations.} This raises the question of whether untracked strategies might affect the interpretation of cooperative and non-cooperative regimes. To address this concern, we complement the profile-based analysis with a state-space perspective that examines occupation times of the four action states $\{(C,C), (C,D), (D,C), (D,D)\}$. 


Because state occupation aggregates behaviour across all possible strategy profiles, this analysis provides a comprehensive and tractable robustness check that incorporates interaction dynamics beyond $\hat{\Pi}_{\epsilon}$. As such, the state-space analysis ensures that our conclusions are informed by the full range of interaction dynamics, including those generated by strategies outside the focal set.

\subsection{Accessible Parameter Range under a Finite Time Horizon}
\label{sec:accessiblepara}

The theoretical analysis in Section~\ref{sec:theory} characterises stable cooperation under the assumption that the learning dynamics converge to a unique stationary distribution in the infinite time limit. In practice, however, computational constraints restrict simulations to a finite time horizon. Throughout our simulations, we therefore fix the horizon at $T_{h} = 200M$ periods.

Before evaluating the quality of the proposed decision boundary, it is necessary to assess under which conditions this finite horizon is sufficient for the learning dynamics to converge to a stationary distribution. Given $T_{h} = 200M$, we define the \emph{accessible parameter range} as the set of algorithmic hyperparameters for which the induced dynamics converge to a stationary distribution within the allotted time horizon.

\subsubsection{Method to Scope the Accessible Parameter Range}

Our approach is to simulate multiple trajectories from diverse initial conditions that span a large portion of the initial Q-value space. If all trajectories converge to the same empirical occupation-time distribution--independently of initialisation--we treat the corresponding parameter configuration as having reached the unique stationary distribution within the chosen time horizon.

\paragraph{Initialisations.}
We consider ten initialisations. These include an optimistic initialisation with $Q_{0}(s,a)=1$ for all $(s,a)$, a pessimistic initialisation with $Q_{0}(s,a)=0$ for all $(s,a)$, and three initialisations at the optimal Q-values corresponding to the Nash equilibrium strategy profiles $AD$, $GT$, and $WSLS$, i.e., $Q_{0}(s,a)=Q^{\text{NE}}(s,a;\theta)$ for all $(s,a)$ and $\text{NE}\in\{\text{AD},\text{GT},\text{WSLS}\}$. In addition, we include five trajectories with Q-values initialised independently and uniformly at random on $[0,1]$.

\paragraph{Stationarity measure.}
To assess convergence across initialisations, we compare empirical occupation times over strategy profiles. Specifically, for a given parameter vector $\theta$, we compute the maximum absolute difference in occupation times across all initialisations and across all strategy profiles in the focal set $\hat{\Pi}_{\epsilon}$, which accounts for the bulk of the stationary distribution:
\begin{equation}
\Delta O(\theta) :=
\max_{i,j\in\mathcal{I}}
\max_{\pi\in \hat{\Pi}_{\epsilon}}
\bigl| O(\tau_{T_{h}}^{i}(\theta), (\pi, \pi))) - O(\tau_{T_{h}}^{j}(\theta), (\pi, \pi))) \bigr|,
\end{equation}
where $\mathcal{I}$ denotes the set of initialisations described above and $\tau_{T_{h}}^{i}(\theta)$ denotes the trajectory generated from initialisation $i$ up to time $T_{h}$ under parameters $\theta$.

We classify a parameter setting $\theta$ as stationary if $\Delta O(\theta) < O^{c}$. In all simulations, we set the threshold to $O^{c}=0.05$.

\paragraph{Illustrative examples.}


\begin{figure}[h!]
    \centering
    \includegraphics[width=1\linewidth]{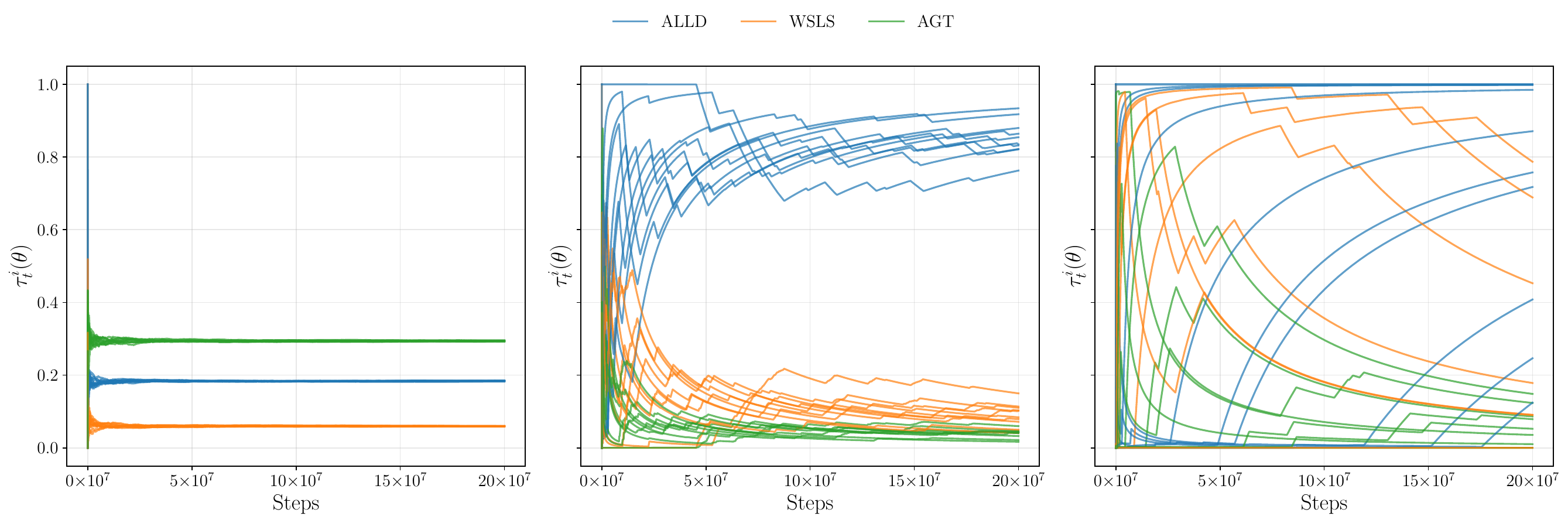}
    \caption{Examples of occupation-time trajectories for ten initialisations under three parameter settings $\theta$. Left: a well-mixed setting in which trajectories converge to the same occupation times within the simulation horizon. Middle: an intermediate case in which trajectories appear close to convergence but exhibit persistent switching across strategy profiles. Right: a non-mixing setting in which occupation times diverge across initialisations even at long horizons. We select strategy profiles that have meaningful shares in at least one of three cases considered.}
    \label{fig:trajectoryexamples}
\end{figure}

Figure~\ref{fig:trajectoryexamples} illustrates the classification procedure. In the left panel, occupation times converge to the same constant across all initialisations within the simulation horizon, indicating convergence to a stationary distribution. In contrast, the right panel shows persistent differences across trajectories, implying that stationarity has not been reached.

The middle panel illustrates a more ambiguous case. While trajectories may appear to approach similar occupation levels, the pronounced switching behaviour suggests slow mixing, with extended periods spent in different strategy profiles. In such cases, empirical occupation times over the considered finite horizon may be misleading. 


\paragraph{Parameter grid.}

From preliminary simulations, we observe that convergence to stationarity depends primarily on the algorithmic parameters $\theta^{a}=(\delta,\epsilon,\alpha)$. For this reason, we fix the environmental parameters $\theta^{e}=(R,P)$ for each $\theta^{a}$ rather than simulating the full five-dimensional parameter space when identifying the accessible parameter range\footnote{Note that we do simulate the full parameter space when assessing the quality of the boundary.}.

Previous work \citep{barfuss2023intrinsic, meylahn2025quantifying} shows that while $AD$ always exists as an equilibrium, $GT$ and $WSLS$ exist only for subsets of the $(R,P)$ space. Given a choice of $\theta^{a}$, we therefore select $R$ and $P$ as the centroid of the region in which all three equilibria coexist.

For $\theta^{a}$, we simulate over the grid $\mathcal{G}=\Theta_{\alpha}^{(m)}\times\Theta_{\epsilon}^{(n)}\times\Theta_{\delta}^{(p)}$, where $\Theta_{\alpha}^{(m)}=\{0.01,0.02,\ldots,0.20\}$, $\Theta_{\epsilon}^{(n)}=\{0.01,0.02,\ldots,0.20\}$, and $\Theta_{\delta}^{(p)}=\{0.65,0.70,\ldots,0.85\}$.

\subsubsection{Results}

\begin{figure}
    \centering
    \includegraphics[width=1.0\linewidth]{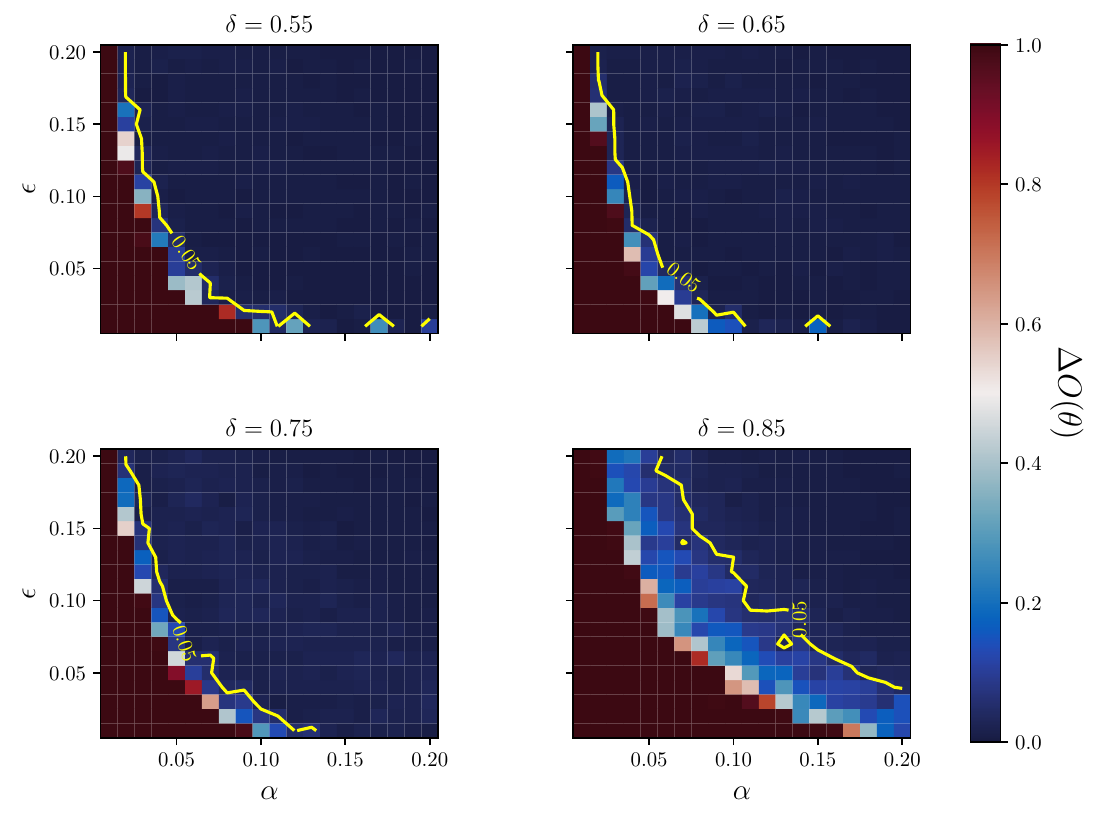}
    \caption{Heatmaps of $\Delta O(\theta)$. Each panel shows values of $\Delta O(\theta)$ across ten initialisations for all combinations of $\alpha$ and $\epsilon$ at a fixed value of $\delta$. Regions with $\Delta O(\theta)<0.05$ are classified as stationary at time $T=2\times10^{8}$.}
    \label{fig:mixing}
\end{figure}

The heatmaps in Figure~\ref{fig:mixing} display $\Delta O(\theta)$ for all combinations of $\alpha\in\Theta_{\alpha}^{(m)}$ and $\epsilon\in\Theta_{\epsilon}^{(n)}$ across different values of $\delta\in\Theta_{\delta}^{(p)}$. Each heatmap also includes the isoquant $\Delta O(\theta)=0.05$. We observe that the boundary separating stationary from non-stationary regions is sharp, relatively smooth, and approximately convex.

The boundary appears to be governed by a nonlinear relationship among learning, exploration, and discounting, consistent with a scaling of the form
\begin{equation}
\epsilon^{c} \sim \frac{\delta}{\alpha}.
\end{equation}

For larger values of $\delta$, such as $\delta=0.85$, the boundary becomes less well defined, with a broad region in which trajectories resembling the intermediate case in Figure~\ref{fig:trajectoryexamples} are common. This observation motivates our choice of $\delta=0.85$ as the upper bound for the discount factor in subsequent simulations. Excluding higher values of $\delta$ is further justified by the fact that the variance of Q-value estimates scales as $(1-\delta)^{-1}$ \citep{devraj2017zap}, implying increasingly large fluctuations as $\delta$ approaches one. Similar effects have been documented for Q-learning in the prisoner's dilemma \citep{denBoer2022artificial}.

Based on these results, we restrict $\alpha$ and $\epsilon$ to values greater than or equal to $0.1$ in the main simulations analysing the heuristic boundary, while retaining the same values of $\delta$. This restriction confines attention to regions of the parameter space in which stationarity is reliably attained, particularly for lower values of $\delta$. For larger $\delta$, the parameter combinations $(\alpha,\epsilon)\in\{(0.1,0.1),(0.1,0.15),(0.15,0.1)\}$ lie close to the boundary between stationary and non-stationary regimes and are therefore treated as boundary cases.

\subsection{Assessing the Quality of the Heuristic Boundary}
\label{sec:assessingboundary}

\paragraph{Parameter grid.}
To assess the quality of the heuristic boundary, we simulate over the grid
\[
\mathcal{G} = \Theta_{\alpha}^{(m)} \times \Theta_{\epsilon}^{(n)} \times \Theta_{\delta}^{(p)} \times \Theta_{P}^{(q)} \times \Theta_{R}^{(s)},
\]
where $\Theta_{\alpha}^{(m)} = \Theta_{\epsilon}^{(n)} = \{0.1, 0.15, 0.2\}$ and $\Theta_{\delta}^{(p)} = \{0.55, 0.65, 0.75, 0.85\}$. For the payoff parameters, we set $\Theta_{P}^{(q)} = \{0.025, 0.05, \ldots, 0.5\}$ and $\Theta_{R}^{(s)} = \{0.525, 0.55, \ldots, 0.975\}$, ensuring that the prisoner's dilemma ordering $T=1>R>P>S=0$ holds throughout. This yields $13{,}680$ parameter combinations. For each combination, we simulate ten trajectories using the initialisations described in Section~\ref{sec:accessiblepara}, resulting in $136{,}000$ trajectories in total. We plot heatmaps of the average empirical occupation time at time $T_{h}$ defined as 
\begin{equation}
\label{eq:ohat}
    \hat{O}(\theta, \boldsymbol{\pi}) = \frac{1}{|\mathcal{I}|} \sum_{i\in\mathcal{I}} O(\tau_{T_{h}}, \boldsymbol{\pi}),
\end{equation}
for the five focal strategy profiles. We normalise by 
\begin{equation}
\label{eq:ohatnorm}
    \hat{O}(\theta, \hat{\Pi}_{\epsilon}):= \sum_{\boldsymbol{\pi}\in \hat{\Pi}_{\epsilon}}\hat{O}(\theta, \boldsymbol{\pi}),
\end{equation}
to make the occupation times comparable across different choices of $\theta$.

\begin{figure}[h!]
    \centering
    \includegraphics[width=1.0\linewidth]{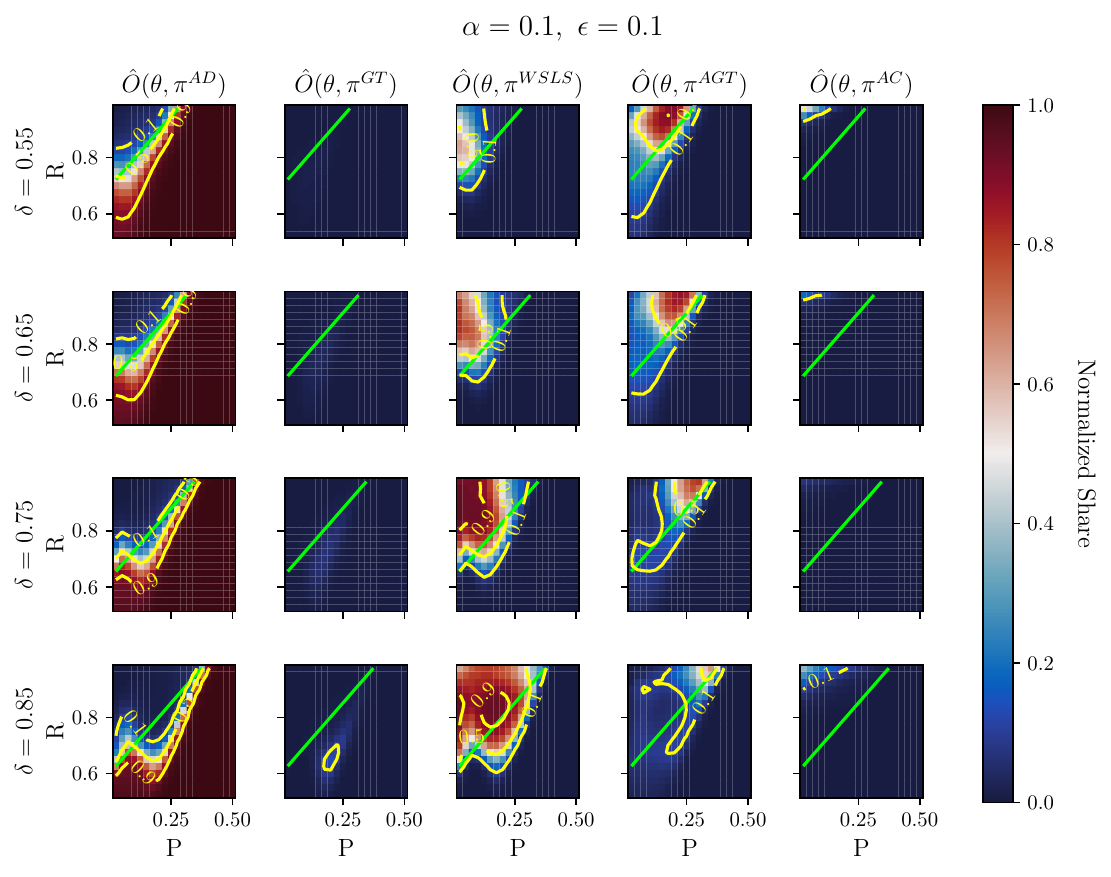}
    \caption{Heatmaps of average empirical occupation times for $AD$, $GT$, $WSLS$, $AGT$, and $AC$ at $\alpha=\epsilon=0.1$. Each heatmap shows the normalised average occupation time across initialisations for all combinations of $P\in\Theta_{P}^{(q)}$ and $R\in\Theta_{R}^{(s)}$, with rows corresponding to different values of $\delta\in\Theta_{\delta}^{(p)}$. Occupation times are normalised by the total time spent in the five focal strategies. The green line denotes the heuristic boundary, while the yellow lines indicate the $10\%$, $50\%$, and $90\%$ isoquants of the normalised occupation shares.}
    \label{fig:summaryplot1}
\end{figure}

Figure~\ref{fig:summaryplot1} provides a visual assessment of the heuristic boundary for the case $\alpha=\epsilon=0.1$. Each heatmap reports the average empirical occupation time of one of the five focal strategy profiles across initialisations, for all combinations of $P\in\Theta_{P}^{(q)}$ and $R\in\Theta_{R}^{(s)}$, and for different values of $\delta\in\Theta_{\delta}^{(p)}$. Occupation times are normalised by the total time spent in the five focal strategies, which allows for consistent comparison of their relative shares within the focal set across parameter configurations


Across the parameter space, the strategy $GT$ is practically absent, a pattern that also holds for other values of $\theta^{a}$. Since the normalised share of cooperative strategies equals one minus the normalised share of non-cooperative strategies--and since $AD$ accounts for the vast majority of non-cooperative play--the quality of the heuristic boundary can be assessed primarily by inspecting the heatmaps for $AD$.


In the $(R,P)$ plane, the heuristic boundary is linear, whereas the empirical boundary inferred from occupation data is nonlinear and deviates increasingly from linearity as $\delta$ increases. This divergence coincides with approaching the non-stationary region of the parameter space. Nevertheless, the heuristic boundary captures the main qualitative features of the empirical transition and, in particular, shifts in the same direction as the empirical boundary when $\epsilon$ varies.


The results for $\alpha=\epsilon=0.1$ shown in Figure~\ref{fig:summaryplot1} are representative of the general pattern observed across parameter combinations. Heatmaps for all remaining cases are reported in Appendix~\ref{app:heatmaps}.

\paragraph{Evaluation Metric.}

To assess the quality of the heuristic boundary in Section~\ref{sec:theory}, we evaluate its predictive performance using standard classification metrics. Interpreting the boundary as a decision rule, Equation~\ref{eq:boundary} yields a prediction of whether cooperation or defection dominates for a given parameter configuration.

We measure predictive performance using the macro F1-score, which jointly captures precision and recall, is invariant to the choice of the positive class, and is robust to class imbalance across regions of the parameter space. The macro F1-score takes values in $[0,1]$, where a value of $1$ corresponds to perfect classification. As a natural benchmark, a trivial classifier that predicts the same outcome for all parameter combinations--such as always predicting defection--achieves a macro F1-score of at most $0.5$. More generally, values close to $0.5$ indicate that the decision rule fails to meaningfully discriminate between cooperative and non-cooperative regimes, while values substantially above $0.5$ reflect increasing discriminatory power. In this sense, macro F1-scores above $0.8$ indicate strong separation between regimes, and scores approaching $1$ indicate near-perfect alignment between the boundary and the empirically observed transition.

If the heuristic boundary is informative, it should therefore achieve macro F1-scores well above this trivial baseline and outperform nearby perturbations of the decision threshold. In particular, relative to small perturbations of the boundary, the heuristic boundary should locally maximise the macro F1-score.

To determine whether cooperation or defection dominates for a given parameter combination, we employ two complementary outcome-labelling criteria. The first criterion is based on time-averaged occupation measures over the five focal strategy profiles described in Section~\ref{sec:simsetup}. The second criterion relies on time-averaged occupation measures over the four joint-action states induced by the actions available to the Q-learning algorithms.




\paragraph{Strategy-profile approach} Under the strategy--profile approach, we label an outcome as \emph{cooperative} if the sum of normalised average occupation times across cooperative strategy profiles exceeds $50\%$. The normalisation divides each profile's raw occupation time by the total time spent in the five focal strategy profiles. We classify $AC$, $WSLS$, and $AGT$ as cooperative, and $AD$ and $GT$ as uncooperative. We classify $GT$ as uncooperative because, in the presence of noise, its stationary distribution concentrates on mutual defection. If our proposed boundary constitutes a meaningful classifier, then using it as a decision rule to predict the dominance of cooperative strategy profiles should yield an F1 score close to one.

Table~\ref{tab:f1_strat} reports the macro F1-scores obtained for different combinations of the learning and exploration rates, pooling observations across the remaining dimensions of the parameter grid $\mathcal{G}$. The table shows excellent and stable F1-scores across the accessible parameter range. 

In Appendix~\ref{app:class_strat}, we report recall and precision scores under an adversarial approach that computes the worst-case recall and precision over definitions of the positive class. Even under this adversarial evaluation, the performance of the boundary remains excellent. Appendix~\ref{app:class_strat} also shows how the performance of the boundary as a decision rule degrades when considering parameter combinations outside the accessible parameter range discussed in Section~\ref{sec:accessiblepara}. 

\begin{table}[t!]
\centering
\begin{minipage}[t]{0.33\textwidth}
\centering
\begin{threeparttable}
\caption{Sum of focal profiles \label{tab:fstrat_sum}}
\input{tables/sum_strat_sum}
\begin{tablenotes}\footnotesize
\item Note: The table reports the mean of the sum of occupation times across the five focal strategy profiles for each combination of the learning and exploration rates. The mean is computed across all hyperparameter settings and payoff specifications within each learning--exploration rate combination. Values in parentheses indicate standard deviations. 
\end{tablenotes}
\end{threeparttable}
\end{minipage}
\hfill
\begin{minipage}[t]{0.6\textwidth}
\centering
\begin{threeparttable}
\caption{F1 Scores \label{tab:f1_strat}}
\input{tables/f1_ep_al_strats}
\begin{tablenotes}\footnotesize
\item Note: The table reports the macro F1 score obtained when using the heuristic boundary as a classifier to predict the dominance of cooperative strategy profiles. Among the five focal strategy profiles, we classify $AC$, $WSLS$, and $AGT$ as cooperative, and $AD$ and $GT$ as uncooperative. The macro F1 score is computed using all hyperparameter settings and payoff specifications within each learning--exploration rate combination. Values in brackets report $95\%$ confidence intervals obtained via a bootstrapping procedure with $1{,}000$ bootstrap samples.
\end{tablenotes}
\end{threeparttable}
\end{minipage}
\end{table}

\begin{figure}[!ht]
    \centering
    \includegraphics[width=1\linewidth]{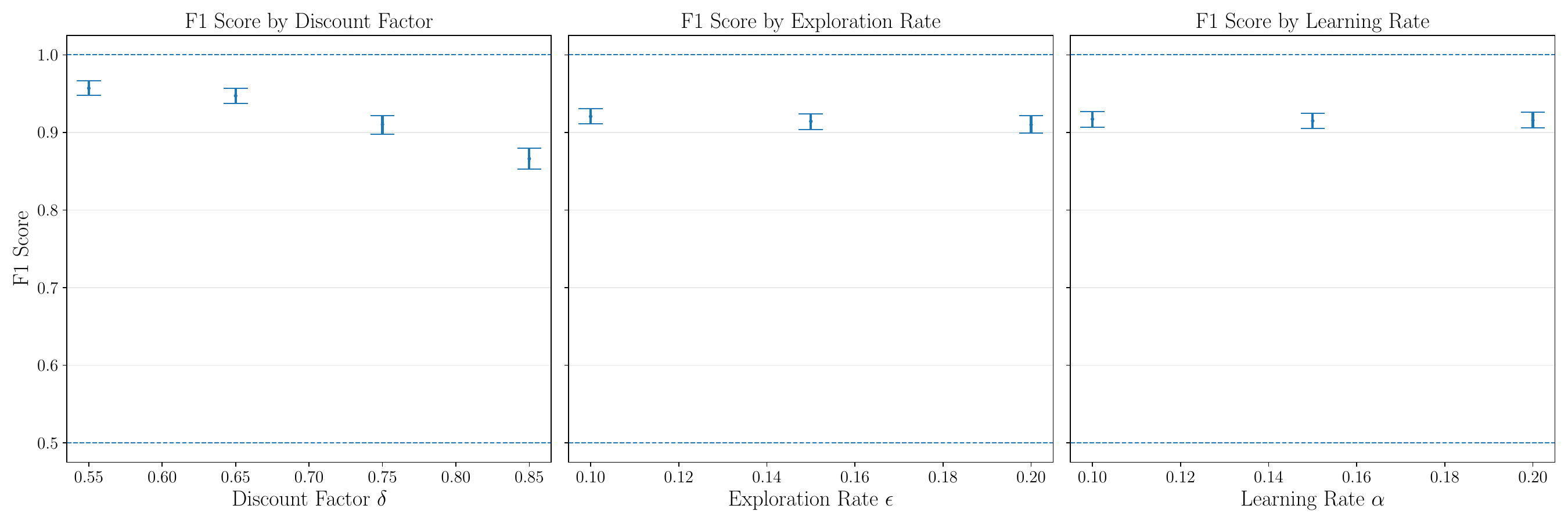}
\caption{Macro F1 scores across algorithmic hyperparameters. For each value of a given hyperparameter, we compute macro F1 scores by aggregating over all remaining hyperparameter combinations and payoff specifications. Error bars indicate $95\%$ confidence intervals obtained via a bootstrapping procedure with $1{,}000$ samples.
\label{fig:f1_strat}}
\end{figure}

Each panel of Figure~\ref{fig:f1_strat} reports the macro F1-score obtained by applying the decision boundary to subsamples of the parameter grid corresponding to different algorithmic hyperparameters. Across all panels, the F1-score is consistently near \(0.9\). The macro F1-score is largely insensitive to the learning rate \(\alpha\) and the exploration rate \(\epsilon\). By contrast, we observe a decline in performance as the discount factor increases. We attribute this effect to the findings discussed in Section~\ref{sec:accessiblepara}, which suggest that higher discount factors push the dynamics toward regimes where the simulation horizon is insufficient for the theoretical framework to apply.

\begin{figure}[h!]
    \centering
    \includegraphics[width=1\linewidth]{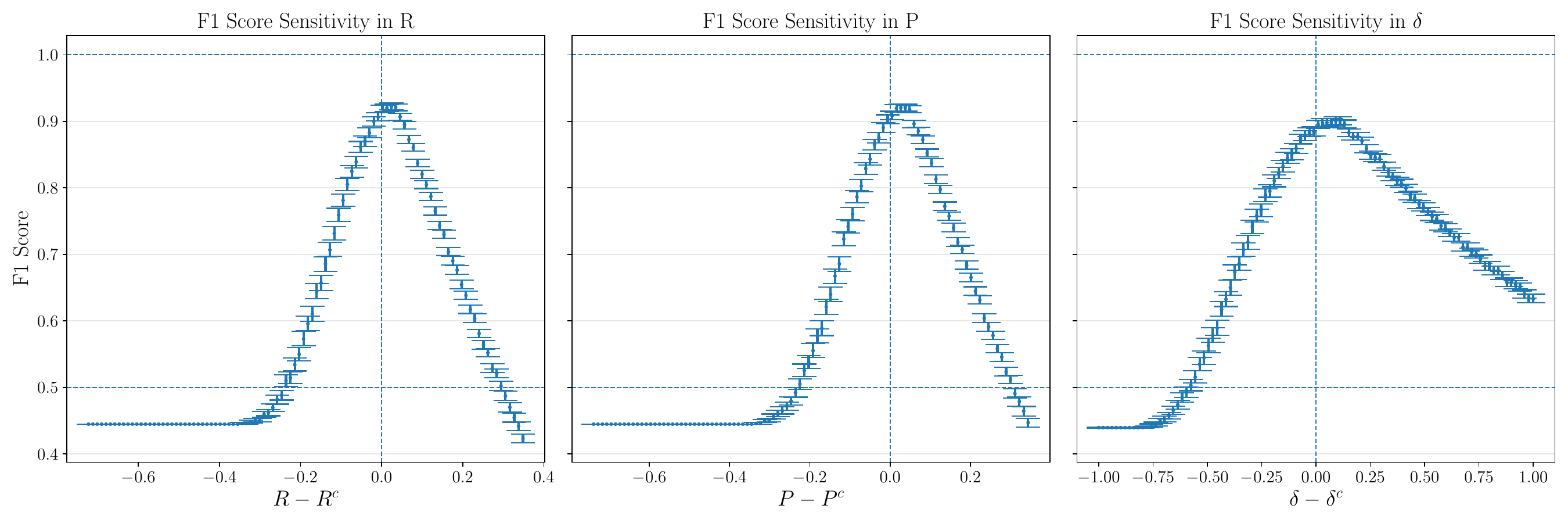}
\caption{Macro F1 scores under perturbations of the boundary in Equation~\eqref{eq:boundary} along different dimensions. Scores are aggregated over all hyperparameter settings and payoff specifications. Error bars indicate $95\%$ confidence intervals from $1{,}000$ bootstrap samples. Perturbations of the discount factor are truncated to $[-1,1]$.
\label{fig:f1_strat_sharpness}}
\end{figure}

Figure~\ref{fig:f1_strat_sharpness} illustrates how the macro F1 score varies as the decision boundary is perturbed along the $R$, $P$, and $\delta$ dimensions. For instance, the left panel replaces the boundary in $R$ implied by Equation~\eqref{eq:boundary}, $R^c$, with an alternative threshold $R'$, and reports the resulting macro F1 score as a function of the Euclidean distance $\lVert R' - R^c \rVert$. Across all dimensions, the decision boundary characterised by Equation~\eqref{eq:boundary} closely aligns with the empirical maximisers of the macro F1 score. Moreover, the macro F1 score decreases monotonically as the boundary deviates from this benchmark, indicating that the proposed boundary is locally optimal and sharply identified.

\paragraph{State-space approach} The state-space approach labels outcomes solely based on the frequency with which the learning dynamics visit the four joint-action states. We thus switch from focusing on $\hat{O}(\theta, \boldsymbol{\pi})$ to $\hat{O}(\theta, s)$, which is defined analogously (see Equations~\eqref{eq:ohat}). This allows us to assess whether the quality of the decision boundary depends on the specific choice of focal strategy profiles or whether it generalises more broadly. Because state occupation times aggregate behaviour across all possible strategy profiles, this approach provides a robustness check that is not tied to the restricted strategy profile set.

\begin{figure}[h!]
    \centering
    \includegraphics[width=1\linewidth]{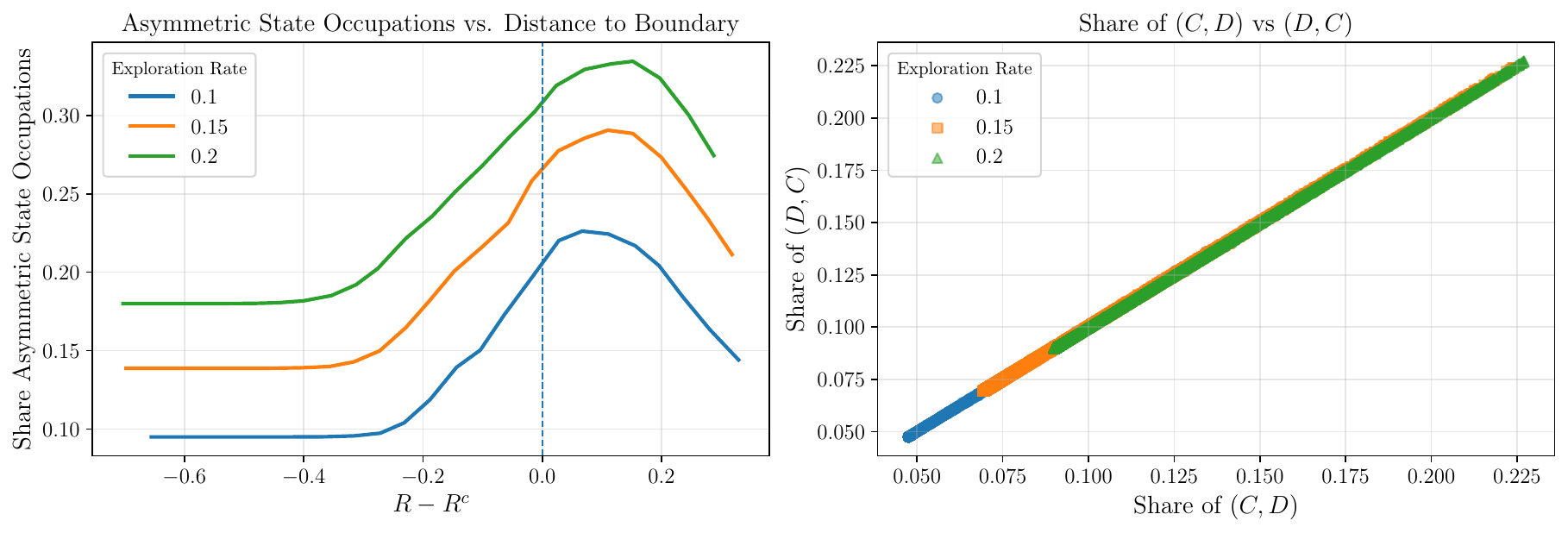}
\caption{Asymmetric state occupations.  The left panel plots the average total occupation time in $(C,D)$ and $(D,C)$ as a function of the distance to the decision boundary along the $R$ dimension, separately for each exploration rate considered in the simulations. The right panel plots, for each simulation run, the share of time spent in state $(D,C)$ against the share of time spent in $(C,D)$, illustrating how players equally split between the roles of traitor and sucker.\label{fig:off_diag}}
\end{figure}

The left panel of Figure~\ref{fig:off_diag} shows that occupation of asymmetric states is most pronounced near the decision boundary and in regions where the boundary predicts cooperation. This pattern suggests that the prevalence of asymmetric profiles--and how they are treated in the definition of cooperation--can materially affect the performance of the proposed boundary.

The right panel of Figure~\ref{fig:off_diag} shows that players split symmetrically between the roles of traitor and sucker. In the parameter regimes considered in our simulations, alternating between the sucker and temptation payoffs yields an average payoff exceeding the Nash equilibrium payoff. This observation provides a rationale for classifying time spent in asymmetric states as cooperative.

\begin{figure}[h!]
    \centering
    \includegraphics[width=1\linewidth]{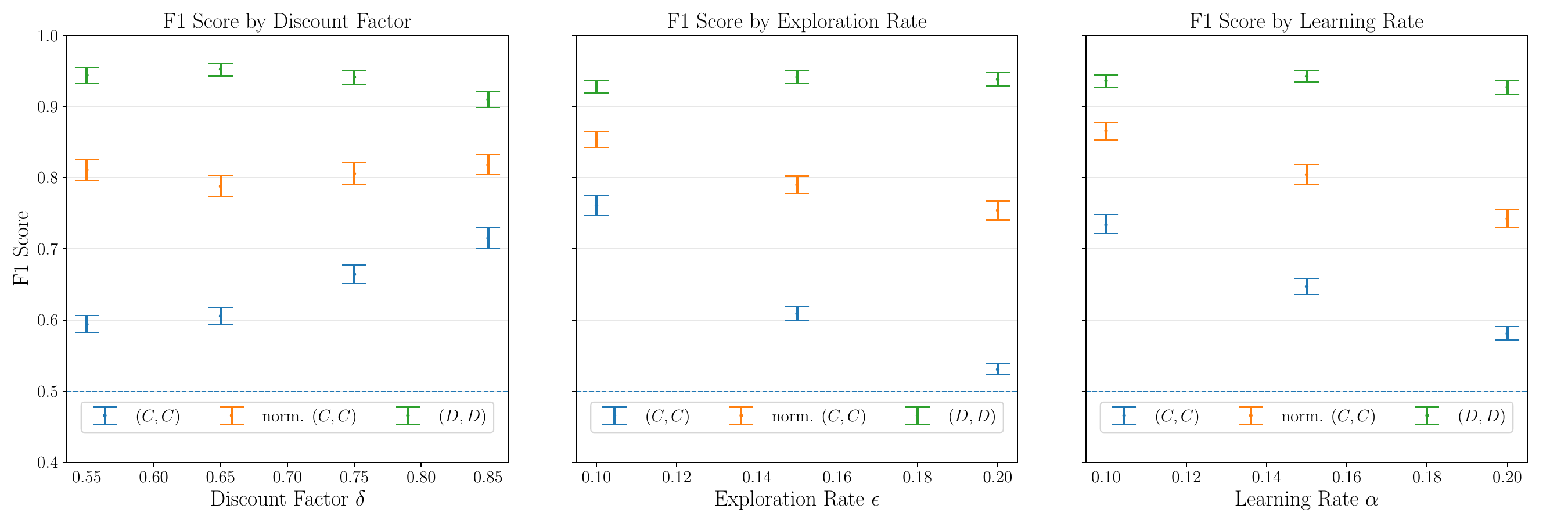}
\caption{Macro F1 scores across algorithmic hyperparameters. For each value of a given hyperparameter, macro F1 scores are computed by aggregating over all remaining hyperparameter combinations and payoff specifications. Error bars indicate $95\%$ confidence intervals obtained via a bootstrapping procedure with $1{,}000$ samples. Scores are reported for three outcome--labelling approaches: dominance of the $(C,C)$ state occupation (blue), dominance of normalised $(C,C)$ occupation (orange), and dominance of the $(D,D)$ state occupation (green).
 \label{fig:f1_states}}
\end{figure}

To avoid relying exclusively on a single definition of cooperation within the state-space approach, we compare three alternative outcome-labelling criteria. The first classifies an outcome as cooperative if the occupation time of state $(D,D)$ is below $50\%$. Although this criterion may be criticised for implicitly treating asymmetric states as cooperative, it is appropriate when the primary objective is to predict whether mutual defection dominates the interaction.

The second approach defines an outcome as cooperative if the occupation time of the state $(C,C)$ is at least $50\%$. While this may appear to be the most natural definition, we anticipate that the prevalence of asymmetric profiles in the region where our boundary predicts cooperation will adversely affect the boundary's performance under this criterion. Moreover, there is a more fundamental limitation to using $(C,C)$ dominance as the benchmark for successful cooperation: the stationary distribution induced by $WSLS$, the strategy profile motivating our heuristic boundary, assigns a non-negligible share of time to $(D,D)$. 

The third approach therefore defines an outcome as cooperative if the occupation time of $(C,C)$ exceeds $50\%$ of the occupation time that the stationary distribution induced by $WSLS$ assigns to $(C,C)$. This definition more closely mirrors the theoretical intuition underlying our boundary, which is motivated by a forgiving cooperative strategy that systematically restores cooperation following unilateral deviations from mutual cooperation.

Figure~\ref{fig:f1_states} reports the macro F1-scores obtained under three alternative outcome-labelling approaches when applying the decision boundary to subsamples of the parameter grid corresponding to different algorithmic hyperparameters. The boundary exhibits strong predictive performance when the objective is to identify the dominance of mutual defection, achieving scores of approximately $0.95$.\footnote{To visualise the quality of the boundary, we report heatmaps of $(D,D)$ state occupation in Appendix~\ref{app:heatmaps_state}.} By contrast, performance deteriorates when predicting the dominance of mutual cooperation, with scores in the range of $0.6$ to $0.7$. 

However, when cooperation rates are normalised relative to the occupation times implied by the stationary distribution induced by $WSLS$, predictive performance improves substantially, with macro F1-scores increasing to between $0.7$ and $0.9$. This improvement relative to the non-normalised definition indicates that the apparent weakness of the boundary under stricter cooperation criteria primarily reflects a mismatch between the outcome metric and the underlying strategic dynamics, rather than a lack of discriminatory power of the boundary itself. 

The remaining performance gap between predicting mutual defection and normalised mutual cooperation reflects a substantive feature of the interaction dynamics: near and above the boundary, asymmetric strategy profiles account for a non-negligible share of the stationary distribution. As a consequence, cooperation criteria that discount asymmetric outcomes tend to conflate regions with qualitatively distinct but closely related dynamics.

Taken together, these results suggest that the boundary is best interpreted as identifying the transition away from defection-dominated behaviour, rather than as a sharp predictor of full mutual cooperation. This interpretation is consistent with the theoretical motivation of the boundary and highlights the importance of outcome definitions that align with the underlying strategic dynamics.

\section{Conclusion}
\label{sec:conclusion}

We have shown and investigated a new form of cooperation between reinforcement learning algorithms. While previous work has looked at the likelihood with which algorithms learn a cooperative strategy, here we instead consider the fraction of time that is spent playing cooperative strategies. This shift in perspective is motivated by the need for algorithms to continue exploring and learning when implemented in practice, due to the possibility of non-stationarities in the environment (or opponent's behaviour). 

Our simulations show that cooperation can also be dominant in this new perspective. Furthermore, we identify a mechanism leading to this domination based on the relative stability of equilibrium strategy profiles and derive a theoretical boundary predicting when cooperation will dominate. This boundary does not take into account all factors relevant for stability, but nevertheless predicts cooperation surprisingly well. 

The phenomenon of spending a majority of the time in cooperative strategy profiles does not yet constitute algorithm collusion, as it would be necessary to argue that the firms are incentivised and likely to use the algorithm (cf. \citep{abada2024algorithmic, den2024mathematical, denBoer2022artificial}). Nevertheless, our results are relevant for the algorithmic collusion literature, as they suggest that algorithmic collusion may not only be a matter of learning collusive strategies, but may require analysis of the robustness of such learnt strategies to fluctuations arising from the environment or from the algorithms' exploration. 

\section{Declaration of generative AI use}
During the preparation of this work, the authors used M365 Copilot and Claude in order to improve the language and phrasing as well as to assist in the coding of the experiments. After using this tool/service, the authors reviewed and edited the content as needed and take full responsibility for the content of the published article.

\bibliographystyle{apalike}
\bibliography{ref.bib}

\newpage

\appendix

\section{Proof of Theorem \ref{thm:recurrentset}}
\label{app:proof}
Note that given any $\boldsymbol{Q}_{t}$, there is a positive probability of both players behaving as if they were playing any strategy in $\Pi_{\epsilon}$ for a finite number of rounds $K$, say $\pi^{i}$ for player $i$. This can be quantified by using, for example, the Kullback-Leibler divergence, but a simple lower bound is given by $\epsilon^{2K}$ since both players explore with probability $\epsilon$. 

In particular, a strategy profile $\boldsymbol{\pi}$ induces a stationary distribution on state-action pairs $\eta^{\boldsymbol{\pi}}(s, a)$. Let 
\begin{equation}
    \kappa_{K}(t) = \{\boldsymbol{a}_{t}, \ldots \boldsymbol{a}_{t+K}\}
\end{equation}
denote a sequence of play of length $K$. Then $\kappa_{K}$ gives rise to an empirical distribution on state-action pairs $\hat{\eta}^{\boldsymbol{\pi}}(\kappa_{K})$ and there exists a set of sequences of play of length $K$ that minimize the total variation distance between the the empirical and true distributions
\begin{equation}
    \kappa_{K}^{\boldsymbol{\pi}} = \argmin_{\kappa_{K}(t)}d_{\text{TV}}(\eta^{\boldsymbol{\pi}}, \hat{\eta}^{\boldsymbol{\pi}}(\kappa_{K}(t))).
\end{equation}
Now at any time $t$, we have 
\begin{align*}
    \mathbb{P}(\kappa_{K}(t)\in\kappa_{K}^{\boldsymbol{\pi}}| \boldsymbol{Q}_{t}) \geq \epsilon^{2K}.
\end{align*}
This is the case even if the strategies of the players, as encoded by their Q-value orderings, change during these $K$ rounds, since there always remains a probability of $\epsilon^{2}$ of the players taking any combination of actions.

From the perspective of a single player, they are facing a stationary environment during these $K$ rounds. For any $\pi^{-i}$ the opponent could be behaving as, it is the case that there is a set of optimal Q-values $Q^{\pi^{-i}}$ encoding the $\text{BR}(\pi^{-i})$. 

This means that we can use the results on the finite-sample complexity of Q-learning with a constant learning rate \citep{beck2012error}. In particular, Theorem 2.1 of \cite{beck2012error} states that, if the $Q_{t}^{i}$ remain bounded for all $i$ and $t$, 
\begin{equation}
\label{eq:beck}
    \mathbb{P}[|Q_{t}^{i} - Q^{\pi^{-i}}|_{\infty}\geq x] \leq \frac{1}{x} \Big((1-\alpha (1-\delta))^{K} + \frac{32\delta}{1-\delta}\sqrt{\frac{\alpha }{2-\alpha}}\Big),
\end{equation}
for any $x>0$. Here, we have used that for us the iterates are bounded by $Q_{t}^{i}\leq Q_{\text{max}}=1$ and $W_{\text{max}} = 4Q_{\text{max}}^{2} |\mathcal{S}\times\mathcal{A}|=32$. 

Given the parameters $\theta$, we can calculate the minimum separation between optimal Q-values over all possible $\pi^{-i}$, which we define as $D(\theta)$, i.e., 
\begin{align}
    D(\theta)=\min_{s\in\mathcal{S},\pi\in\Pi_{\epsilon}}|Q^{\pi}(s, C) - Q^{\pi}(s, D)|.
\end{align}

From Equation~\eqref{eq:beck} we obtain that there is a positive probability of all Q-values of player $i$ being within $D(\theta)/2$ of $Q^{\pi^{-i}}$ within $K$ rounds for some $K$ and $\alpha<\alpha^{c}$, for some $\alpha^{c}$. In particular,
\begin{equation}
    \mathbb{P}\Big[|Q_{t}^{i} - Q^{\pi^{-i}}|_{\infty}<\frac{D(\theta)}{2}\Big] \geq 1-\frac{2}{D(\theta)} \Big((1-\alpha (1-\delta))^{K} + \frac{32\delta}{1-\delta}\sqrt{\frac{\alpha }{2-\alpha}}\Big).
\end{equation}
Thus the event $|Q_{t}^{i} - Q^{\pi^{-i}}|_{\infty}<\frac{D(\theta)}{2}$ occurs with positive probability if
\begin{equation}
    \frac{2}{D(\theta)} \Big((1-\alpha (1-\delta))^{K} + \frac{32\delta}{1-\delta}\sqrt{\frac{\alpha }{2-\alpha}}\Big)<1.
\end{equation}
The term $(1-\alpha (1-\delta))^{K}$ can be made arbitrarily small by increasing $K$ since $\alpha, \delta\in(0, 1)$. This means that a large enough $K$ to satisfy the inequality exists if 
\begin{equation}
    \frac{2}{D(\theta)} \frac{32\delta}{1-\delta}\sqrt{\frac{\alpha }{2-\alpha}} < 1.
\end{equation}
Solving for $\alpha$, gives that such a $K$ exists if 
\begin{equation}
    \alpha < \alpha_{c}:=\frac{2D^{2}(\theta)(1-\delta)^{2}}{D^{2}(\theta)(1-\delta)^{2} + 4096\,\delta^{2}}.
\end{equation}

This brings the Q-values of a player close enough to the optimal Q-values to guarantee that they will have transitioned to the best-response strategy. Since this holds for both players, there is a positive probability that after $K$ rounds the players are playing $(\text{BR}(\pi^{-1}), \text{BR}(\pi^{1}))$.

We conclude that there is a positive probability of transitioning from any strategy profile to a strategy profile where both players are playing the best response to some strategy within $K$ rounds when $K$ is large enough. The set of strategy profiles
\begin{align}
   \Pi_{a} := \{(\tilde{\pi}^{1}, \tilde{\pi}^{-1}) : \tilde{\pi}^{i}\in \text{BR}(\pi) \text{ for some } \pi\in\Pi_{\epsilon}\}
\end{align}
is thus accessible from all $\boldsymbol{\pi}$ and thus forms part of a communicating class $\Pi_{c}$ if $\alpha<\alpha_{c}$. 

To obtain that they thus form part of a unique recurrent set, note that if $\alpha<\alpha_{c}$, we can pick $K$ such that the probability $\zeta$ of hitting $\Pi_{c}$ is bounded away from zero for any state the process could be in at the beginning of the $K$ time periods. Splitting the time horizon into non-overlapping blocks of length $K$ and applying the result in each block, we obtain that the strategy profiles in $\Pi_{a}$ will be visited infinitely often. This completes the proof.

\section{Details of heuristic boundary derivation}
\label{app:criticalboundary}
In this appendix, we provide the details of the derivation of the critical boundary. By solving the Bellman optimality equation, assuming that both players play the AD strategy, we find that the optimal Q-values are
\begin{align}
    Q^{\text{AD}}(s, D;\theta) = \frac{1}{2}\left(\epsilon (T-P) + 2P \right),
\end{align}
and
\begin{align}
    Q^{\text{AD}}(s, C) = \frac{1}{2}\left( 2 \delta P - \delta \epsilon P + \epsilon R - \delta \epsilon R + 2 S - 2 \delta S -  \epsilon S + \delta \epsilon S + \delta \epsilon T\right),
\end{align}
for all $s\in\mathcal{S}$. Similarly, 
\begin{align}
    Q^{\text{WSLS}}(s, D;\theta) = &\frac{1}{4}[4T +(2 \delta (1 - \epsilon) + \epsilon) \{2 P + \delta (2 - \epsilon) (R-P) + \delta \epsilon (S - T) - 2 T\}]
\end{align}
and 
\begin{align}
    Q^{\text{WSLS}}(s, C;\theta) = \frac{1}{4}[2 (2 - \epsilon) R + 2 \epsilon S + \delta \epsilon \{(2 - \epsilon) P - 2 R + \epsilon (R - S + T)\}]
\end{align}
for $s\in\{(D, D), (C, C)\}$, and
\begin{align}
    Q^{\text{WSLS}}(s, D;\theta) = \frac{1}{4}[(2 - \delta (2 - \epsilon)) (2 - \epsilon) P + 2 \epsilon T - \delta (2 - \epsilon) \{\epsilon (T-S) - (2 - \epsilon) R\}]
\end{align}
and
\begin{align}
     Q^{\text{WSLS}}(s, C;\theta) = \frac{1}{4}[2 \epsilon (R - S) + 4 S + 2 \delta^2 (1 - \epsilon) \{2R - (2 - \epsilon) P - \epsilon (R - S + T)\}\nonumber \\
     +\delta \{(2 - \epsilon)^2 P - 4 S + \epsilon (2 (S + T)-\epsilon (R - S + T) )\}]
\end{align}
for $s\in\{(D, C), (C, D)\}$. Since the Q-values for the two actions are the same in all states when considering AD, we find that 
\begin{align}
\label{eq:minAD}
    \min_{s\in\mathcal{S}}\Delta Q^{\text{AD}}(s;\theta) = \frac{1}{2} [(2 - \epsilon) P - 2 S + \epsilon (S + T-R)].
\end{align}
For WSLS, the states are split into two pairs in terms of their Q-value difference so 
\begin{align*}
    \Delta Q^{\text{WSLS}}(s ;\theta) = R - T + \frac{\epsilon}{2} (T+S - (P + R)) + \frac{\delta}{2} (1 - \epsilon) \{2R-(2 - \epsilon) P - \epsilon (T+R-S)\}
\end{align*}
for $s\in\{(D, D), (C, C)\}$ and
\begin{align*}
    \Delta Q^{\text{WSLS}}(s ;\theta) = \frac{1}{2} [(1 - \delta (1 - \epsilon)) (2 - \epsilon) P - 2 S + \epsilon (T + S - R) + \delta (1 - \epsilon) \{(2 - \epsilon) R - \epsilon (T-S)\}]
\end{align*}
for $s\in\{(D, C), (C, D)\}$. Now we have
\begin{align}
    \Delta Q^{\text{WSLS}}((D, C) ;\theta) - \Delta Q^{\text{WSLS}}((D, D) ;\theta) = T-R + P - S > 0
\end{align}
so that
\begin{align}
\Delta Q^{\text{WSLS}}((D, D) ;\theta) = \Delta Q^{\text{WSLS}}((C, C) ;\theta) < \Delta Q^{\text{WSLS}}((D, C) ;\theta) = \Delta Q^{\text{WSLS}}((C, D) ;\theta).
\end{align}
Therefore
\begin{align}
\label{eq:minWSLS}
    \min_{s\in\mathcal{S}}\Delta Q^{\text{WSLS}}(s;\theta) =R - T + \frac{\epsilon}{2} (T+S - (P + R)) + \frac{\delta}{2} (1 - \epsilon) \{2R-(2 - \epsilon) P - \epsilon (T+R-S)\}.
\end{align}
By equating Equation~\eqref{eq:minAD} and Equation~\eqref{eq:minWSLS} and solving for $\delta$, we obtain the boundary. 
\clearpage

\section{Normalized All-Defection Heat Maps}
\label{app:heatmaps}

In this appendix, we present heatmaps of the average empirical occupation time of $AD$ across all simulated pairs of the learning rate $\alpha$ and the exploration rate $\epsilon$. These heatmaps visualize how frequently the interaction converges to persistent defection as algorithmic parameters vary. We focus on $AD$ because, among the five focal strategy profiles, it is the only non-cooperative strategy that attains substantial occupation time in our simulations; by contrast, $GT$ is almost never observed. As a result, occupation of $AD$ serves as a clear and informative proxy for the failure of cooperation and provides a succinct evaluation of the effectiveness of the proposed decision boundary.


\begin{figure}[ht]
    \centering
    \includegraphics[width=1.0\linewidth]{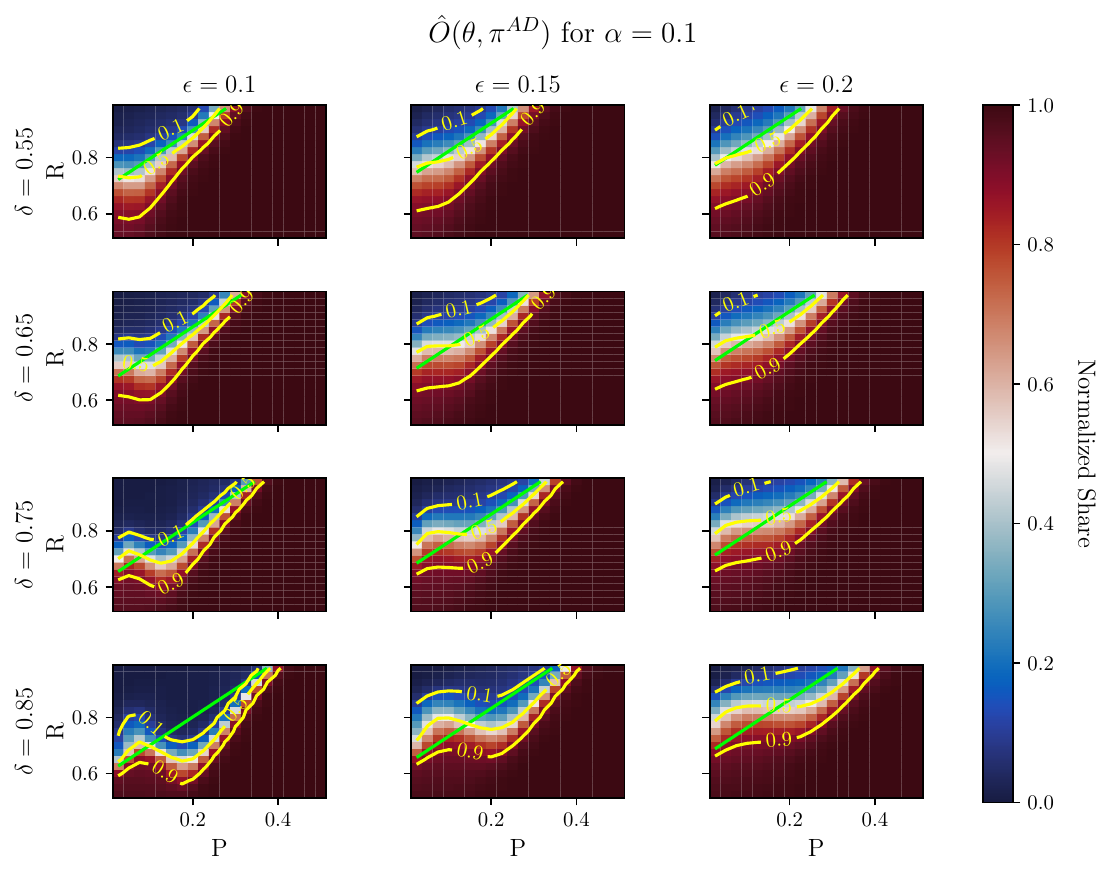}
    \caption{Heatmaps of the average empirical occupation time of $AD$ at $\alpha = 0.1$. Each heatmap reports the normalized average occupation time, averaged across initializations, for all combinations of $P \in \Theta_{P}^{(q)}$ and $R \in \Theta_{R}^{(s)}$. Rows correspond to different values of $\delta \in \Theta_{\delta}^{(p)}$, and columns correspond to different values of $\epsilon \in \Theta_{\epsilon}^{(n)}$. Raw occupation times are normalized by dividing by the total occupation time across the five focal strategy profiles ($AD$, $GT$, $WSLS$, $AGT$, and $AC$). The green line indicates the heuristic boundary, while the yellow lines denote the $10\%$, $50\%$, and $90\%$ isoquants of the normalized occupation share. Regions with high $AD$ occupation coincide closely with the boundary's prediction of non-cooperative outcomes.}\label{fig:a1}
\end{figure}


\begin{figure}[ht]
    \centering
    \includegraphics[width=1.0\linewidth]{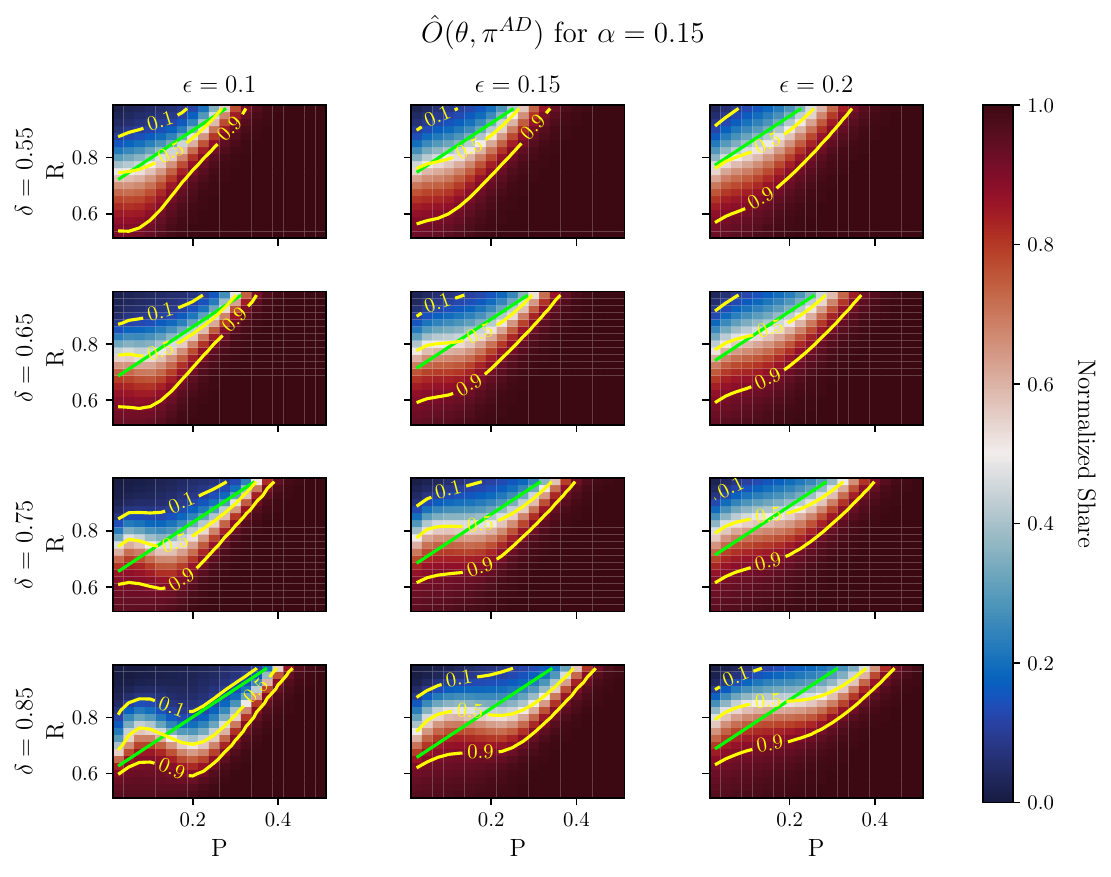}
    \caption{Heatmaps of the average empirical occupation time of $AD$ at $\alpha = 0.15$. Each heatmap reports the normalized average occupation time, averaged across initializations, for all combinations of $P \in \Theta_{P}^{(q)}$ and $R \in \Theta_{R}^{(s)}$. Rows correspond to different values of $\delta \in \Theta_{\delta}^{(p)}$, and columns correspond to different values of $\epsilon \in \Theta_{\epsilon}^{(n)}$. Raw occupation times are normalized by dividing by the total occupation time across the five focal strategy profiles ($AD$, $GT$, $WSLS$, $AGT$, and $AC$). The green line indicates the heuristic boundary, while the yellow lines denote the $10\%$, $50\%$, and $90\%$ isoquants of the normalized occupation share. Regions with high $AD$ occupation coincide closely with the boundary's prediction of non-cooperative outcomes.}\label{fig:a15}
\end{figure}


\begin{figure}[ht]
    \centering
    \includegraphics[width=1.0\linewidth]{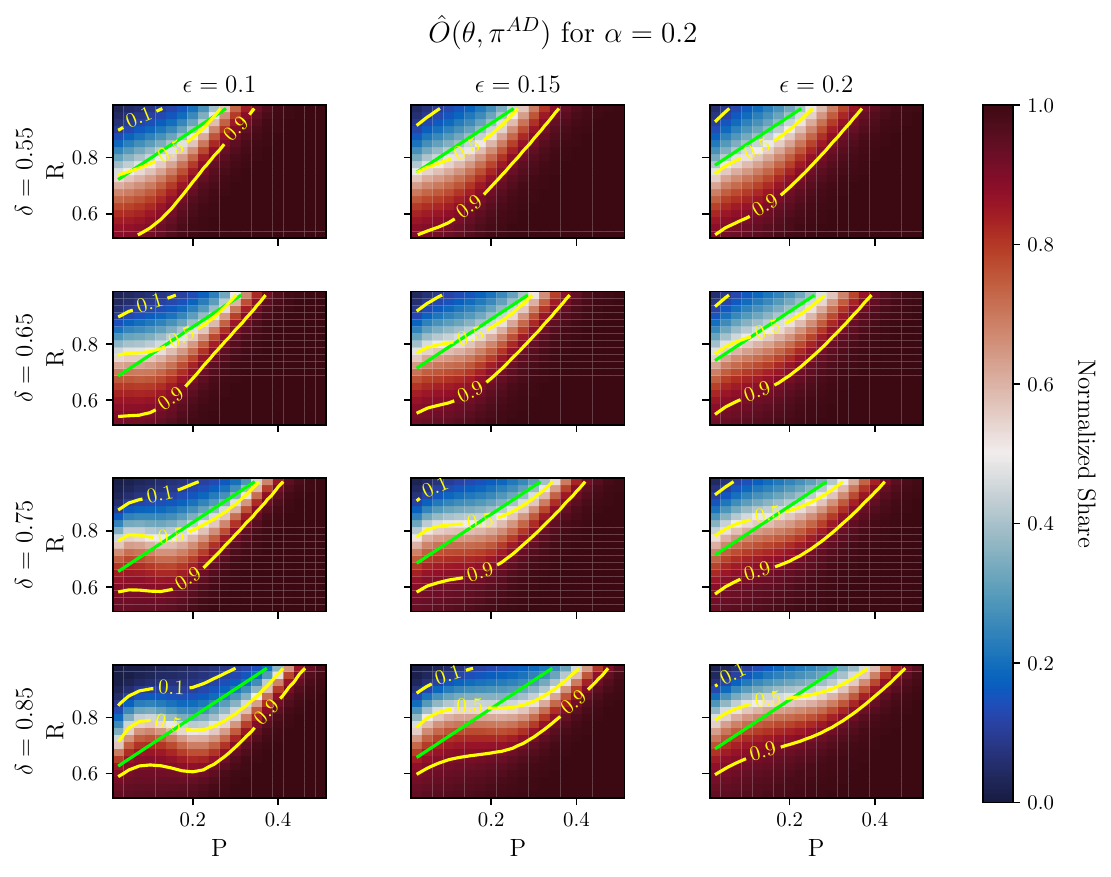}
     \caption{Heatmaps of the average empirical occupation time of $AD$ at $\alpha = 0.2$. Each heatmap reports the normalized average occupation time, averaged across initializations, for all combinations of $P \in \Theta_{P}^{(q)}$ and $R \in \Theta_{R}^{(s)}$. Rows correspond to different values of $\delta \in \Theta_{\delta}^{(p)}$, and columns correspond to different values of $\epsilon \in \Theta_{\epsilon}^{(n)}$. Raw occupation times are normalized by dividing by the total occupation time across the five focal strategy profiles ($AD$, $GT$, $WSLS$, $AGT$, and $AC$). The green line indicates the heuristic boundary, while the yellow lines denote the $10\%$, $50\%$, and $90\%$ isoquants of the normalized occupation share. Regions with high $AD$ occupation coincide closely with the boundary's prediction of non-cooperative outcomes.}\label{fig:a2}
\end{figure}

\clearpage

\section{State Occupation Heat Maps for $(D,D)$}
\label{app:heatmaps_state}

In this appendix, we present heatmaps of the average empirical occupation time of the state $(D,D)$ across all simulated combinations of the learning rate $\alpha$ and the exploration rate $\epsilon$. These heatmaps illustrate how frequently the state $(D,D)$ is visited as the algorithmic parameters vary. In contrast to the analysis of strategy profiles, for which we report normalized shares, we use raw (unnormalized) shares in this case.


\begin{figure}[ht]
    \centering
    \includegraphics[width=1.0\linewidth]{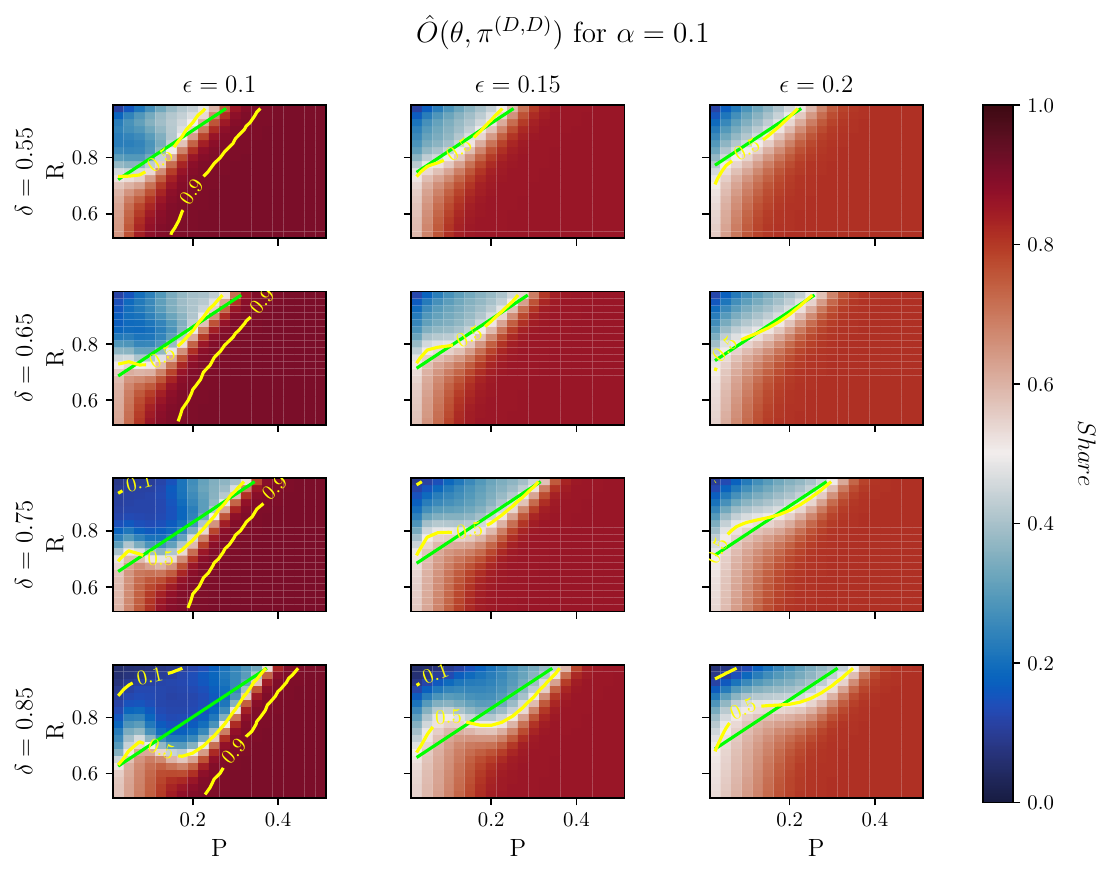}
    \caption{Heatmaps of the average empirical occupation time of $(D,D)$ at $\alpha = 0.1$. Each heatmap reports the normalized average occupation time, averaged across initializations, for all combinations of $P \in \Theta_{P}^{(q)}$ and $R \in \Theta_{R}^{(s)}$. Rows correspond to different values of $\delta \in \Theta_{\delta}^{(p)}$, and columns correspond to different values of $\epsilon \in \Theta_{\epsilon}^{(n)}$. The green line indicates the heuristic boundary, while the yellow lines denote the $10\%$, $50\%$, and $90\%$ isoquants.}
\end{figure}


\begin{figure}[ht]
    \centering
    \includegraphics[width=1.0\linewidth]{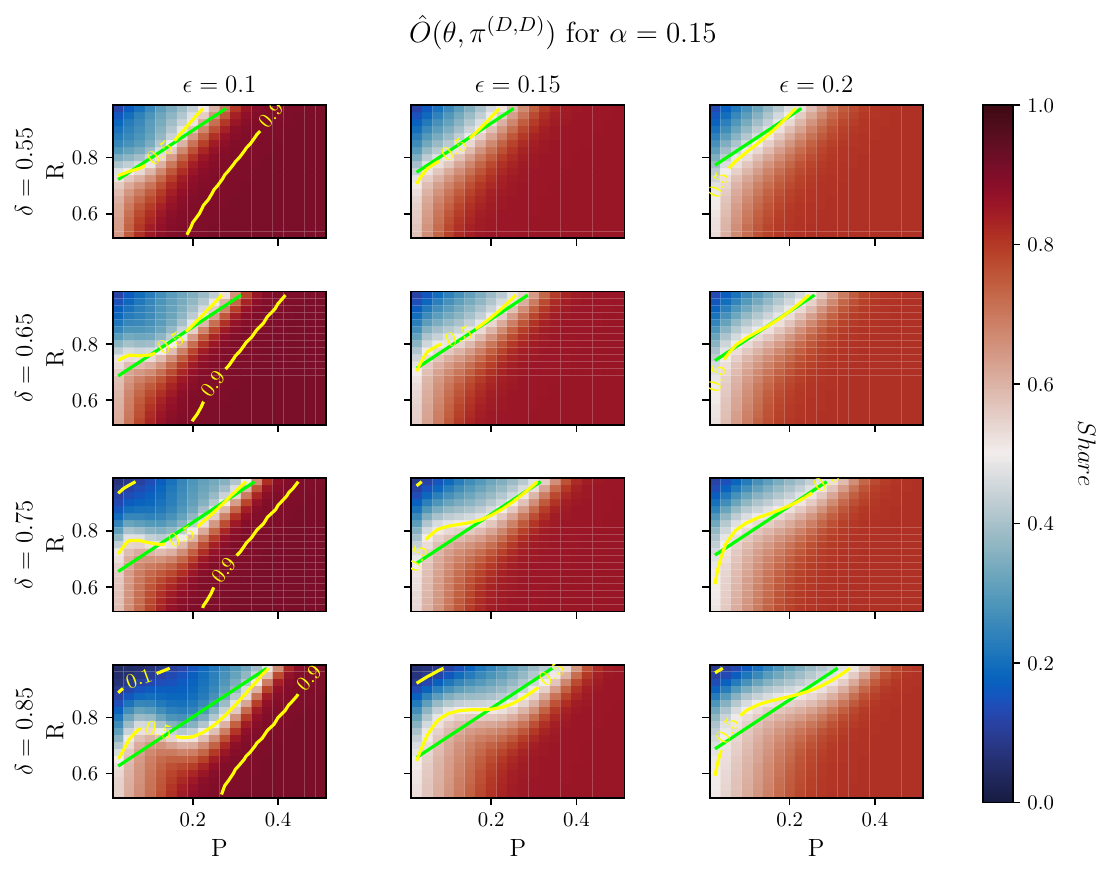}
    \caption{Heatmaps of the average empirical occupation time of $(D,D)$ at $\alpha = 0.15$. Each heatmap reports the normalized average occupation time, averaged across initializations, for all combinations of $P \in \Theta_{P}^{(q)}$ and $R \in \Theta_{R}^{(s)}$. Rows correspond to different values of $\delta \in \Theta_{\delta}^{(p)}$, and columns correspond to different values of $\epsilon \in \Theta_{\epsilon}^{(n)}$. The green line indicates the heuristic boundary, while the yellow lines denote the $10\%$, $50\%$, and $90\%$ isoquants.}
\end{figure}


\begin{figure}[ht]
    \centering
    \includegraphics[width=1.0\linewidth]{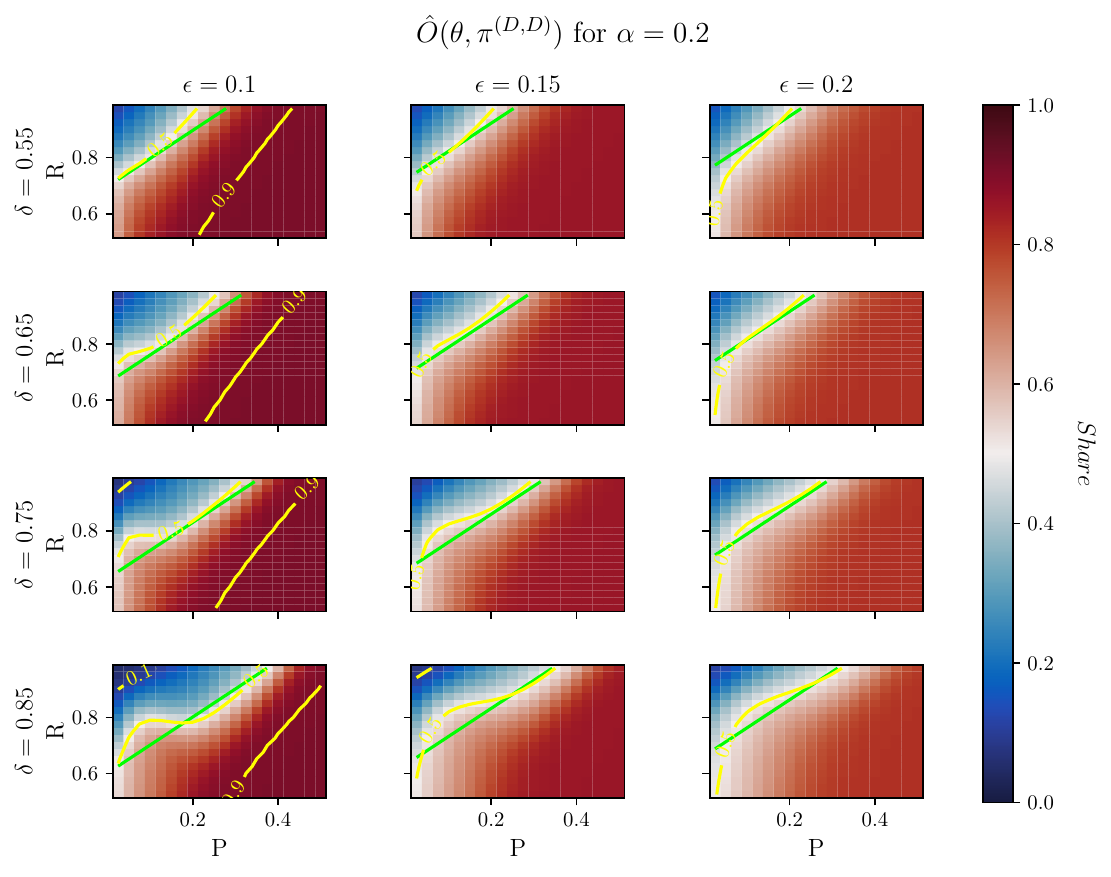}
     \caption{Heatmaps of the average empirical occupation time of $AD$ at $\alpha = 0.2$. Each heatmap reports the normalized average occupation time, averaged across initializations, for all combinations of $P \in \Theta_{P}^{(q)}$ and $R \in \Theta_{R}^{(s)}$. Rows correspond to different values of $\delta \in \Theta_{\delta}^{(p)}$, and columns correspond to different values of $\epsilon \in \Theta_{\epsilon}^{(n)}$. The green line indicates the heuristic boundary, while the yellow lines denote the $10\%$, $50\%$, and $90\%$ isoquants.}
\end{figure}

\clearpage

\section{F1 Scores, Precision and Recall for Dominance of Cooperative Strategies}
\label{app:class_strat}

This appendix reports the macro F1 score, minimum precision, and minimum recall across combinations of the learning and exploration rates. For completeness, we also present results for two combinations of the learning and exploration rates that lie outside the accessible parameter range identified in Section~\ref{sec:accessiblepara}, namely $\alpha = \epsilon = 0.025$ and $\alpha = \epsilon = 0.05$. For these parameter combinations, we evaluate the classifier over the same values of the remaining free parameters as in the original grid $\mathcal{G}$.

Table~\ref{tab:f1_strat_tot} reports the macro F1 scores. These results replicate those in Table~\ref{tab:f1_strat} while extending the analysis to the non-accessible parameter range. The table reveals a monotonic decline in the macro F1 score as the learning and exploration rates move further into the non-accessible region.

A potential limitation of the macro F1 score is that it may mask poor performance in either recall or precision, depending on the definition of the positive class (dominance of cooperative versus defection strategy profiles). To address this concern, Tables~\ref{tab:pr_strat} and~\ref{tab:rec_strat} adopt an adversarial, worst-case evaluation approach. Specifically, for each parameter combination, we compute recall and precision under both possible definitions of the positive class and report the minimum value. Both metrics range from $0$ to $1$, with lower values indicating weaker performance.

The results show strong performance across all metrics within the accessible parameter range, but substantially degraded performance outside this range.

\begin{table}[ht!]
\begin{threeparttable}
\caption{F1 Scores \label{tab:f1_strat_tot}}
\input{tables/f1_ep_al_strats_tot}
\begin{tablenotes}\footnotesize
\item \emph{Note:} The table reports the macro F1 score obtained when using the heuristic boundary as a classifier to predict the dominance of cooperative strategy profiles. Values in brackets indicate $95\%$ confidence intervals obtained via a bootstrapping procedure with $1{,}000$ bootstrap samples. Performance degrades for parameter combinations outside the accessible range ($\alpha = \epsilon = 0.025$ and $\alpha = \epsilon = 0.05$). For these combinations of the learning and exploration rates, the classifier is evaluated over the remaining free parameters of the original grid, $\Theta_{\delta}^{(p)} \times \Theta_{P}^{(q)} \times \Theta_{R}^{(s)}$.
\end{tablenotes}
\end{threeparttable}
\end{table}

\begin{table}[ht!]
\begin{threeparttable}
\caption{Minimum Precision Scores \label{tab:pr_strat}}
\input{tables/pr_ep_al_strats}
\begin{tablenotes}\footnotesize
\item \emph{Note:} The table reports the minimum precision obtained when using the heuristic boundary as a classifier to predict the dominance of cooperative strategy profiles. Values in brackets indicate $95\%$ confidence intervals obtained via a bootstrapping procedure with $1{,}000$ bootstrap samples. With two classes, precision is computed for each class and the minimum value is reported.
\end{tablenotes}
\end{threeparttable}
\end{table}

\begin{table}[ht!]
\begin{threeparttable}
\caption{Minimum Recall Scores\label{tab:rec_strat}}
\input{tables/rec_ep_al_strats}
\begin{tablenotes}\footnotesize
\item \emph{Note:} The table reports the minimum recall obtained when using the heuristic boundary as a classifier to predict the dominance of cooperative strategy profiles. Values in brackets indicate $95\%$ confidence intervals obtained via a bootstrapping procedure with $1{,}000$ bootstrap samples. With two classes, recall is computed for each class and the minimum value is reported.
\end{tablenotes}
\end{threeparttable}
\end{table}


\end{document}

%% file: tables/sum_strat_sum.tex
\begin{tabular}{lccc}
\toprule
\diagbox[width=0.75cm,height=0.75cm]{$\epsilon$}{$\alpha$} & 0.1 & 0.15 & 0.2 \\
\hline
\multirow{2}{*}{0.1} & 0.901 & 0.837 & 0.814 \\
 & (0.144) & (0.200) & (0.225) \\
\bottomrule
\multirow{2}{*}{0.15} & 0.851 & 0.810 & 0.792 \\
 & (0.199) & (0.235) & (0.246) \\
\bottomrule
\multirow{2}{*}{0.2} & 0.825 & 0.791 & 0.771 \\
 & (0.232) & (0.254) & (0.260) \\
\bottomrule
\end{tabular}

%% file: tables/f1_ep_al_strats.tex
\begin{tabular}{lccc}
\toprule
\diagbox[width=0.75cm,height=0.75cm]{$\epsilon$}{$\alpha$} & 0.1 & 0.15 & 0.2 \\
\hline
\multirow{2}{*}{0.1} & 0.918 & 0.924 & 0.920 \\
 & [0.902, 0.933] & [0.907, 0.940] & [0.904, 0.936] \\
\bottomrule
\multirow{2}{*}{0.15} & 0.921 & 0.910 & 0.912 \\
 & [0.904, 0.938] & [0.891, 0.929] & [0.893, 0.929] \\
\bottomrule
\multirow{2}{*}{0.2} & 0.908 & 0.908 & 0.914 \\
 & [0.888, 0.927] & [0.888, 0.926] & [0.894, 0.931] \\
\bottomrule
\end{tabular}

%% file: tables/f1_ep_al_strats_tot.tex
\begin{tabular}{lccccc}
\toprule
\diagbox[width=0.75cm,height=0.75cm]{$\epsilon$}{$\alpha$} & 0.025 & 0.05 & 0.1 & 0.15 & 0.2 \\
\hline
\multirow{2}{*}{0.025} & 0.819 & -- & -- & -- & -- \\
 & [0.799, 0.838] & -- & -- & -- & -- \\
\bottomrule
\multirow{2}{*}{0.05} & -- & 0.849 & -- & -- & -- \\
 & -- & [0.831, 0.867] & -- & -- & -- \\
\bottomrule
\multirow{2}{*}{0.1} & -- & -- & 0.918 & 0.924 & 0.920 \\
 & -- & -- & [0.902, 0.933] & [0.907, 0.940] & [0.904, 0.936] \\
\bottomrule
\multirow{2}{*}{0.15} & -- & -- & 0.921 & 0.910 & 0.912 \\
 & -- & -- & [0.904, 0.938] & [0.891, 0.929] & [0.893, 0.929] \\
\bottomrule
\multirow{2}{*}{0.2} & -- & -- & 0.908 & 0.908 & 0.914 \\
 & -- & -- & [0.888, 0.927] & [0.888, 0.926] & [0.894, 0.931] \\
\bottomrule
\end{tabular}

%% file: tables/pr_ep_al_strats.tex
\begin{tabular}{lccccc}
\toprule
\diagbox[width=0.75cm,height=0.75cm]{$\epsilon$}{$\alpha$} & 0.025 & 0.05 & 0.1 & 0.15 & 0.2 \\
\hline
\multirow{2}{*}{0.025} & 0.773 & -- & -- & -- & -- \\
 & [0.755, 0.791] & -- & -- & -- & -- \\
\bottomrule
\multirow{2}{*}{0.05} & -- & 0.817 & -- & -- & -- \\
 & -- & [0.801, 0.834] & -- & -- & -- \\
\bottomrule
\multirow{2}{*}{0.1} & -- & -- & 0.933 & 0.918 & 0.879 \\
 & -- & -- & [0.922, 0.945] & [0.889, 0.945] & [0.847, 0.912] \\
\bottomrule
\multirow{2}{*}{0.15} & -- & -- & 0.907 & 0.855 & 0.845 \\
 & -- & -- & [0.875, 0.940] & [0.818, 0.896] & [0.807, 0.884] \\
\bottomrule
\multirow{2}{*}{0.2} & -- & -- & 0.844 & 0.836 & 0.844 \\
 & -- & -- & [0.799, 0.887] & [0.794, 0.879] & [0.801, 0.884] \\
\bottomrule
\end{tabular}

%% file: tables/rec_ep_al_strats.tex
\begin{tabular}{lccccc}
\toprule
\diagbox[width=0.75cm,height=0.75cm]{$\epsilon$}{$\alpha$} & 0.025 & 0.05 & 0.1 & 0.15 & 0.2 \\
\hline
\multirow{2}{*}{0.025} & 0.636 & -- & -- & -- & -- \\
 & [0.598, 0.672] & -- & -- & -- & -- \\
\bottomrule
\multirow{2}{*}{0.05} & -- & 0.668 & -- & -- & -- \\
 & -- & [0.632, 0.705] & -- & -- & -- \\
\bottomrule
\multirow{2}{*}{0.1} & -- & -- & 0.805 & 0.850 & 0.872 \\
 & -- & -- & [0.768, 0.842] & [0.815, 0.886] & [0.837, 0.904] \\
\bottomrule
\multirow{2}{*}{0.15} & -- & -- & 0.843 & 0.852 & 0.869 \\
 & -- & -- & [0.804, 0.881] & [0.811, 0.893] & [0.830, 0.908] \\
\bottomrule
\multirow{2}{*}{0.2} & -- & -- & 0.848 & 0.854 & 0.866 \\
 & -- & -- & [0.802, 0.893] & [0.808, 0.895] & [0.819, 0.908] \\
\bottomrule
\end{tabular}

%% file: ref.bib
@article{beck2012error,
  title={Error bounds for constant step-size {Q}-learning},
  author={Beck, Carolyn L and Srikant, Rayadurgam},
  journal={Systems \& Control Letters},
  volume={61},
  number={12},
  pages={1203--1208},
  year={2012},
  publisher={Elsevier}
}

@article{kushner1981asymptotic,
  title={Asymptotic properties of stochastic approximations with constant coefficients},
  author={Kushner, Harold J and Huang, Hai},
  journal={SIAM Journal on Control and Optimization},
  volume={19},
  number={1},
  pages={87--105},
  year={1981},
  publisher={SIAM}
}

@article{kushner1981averaging,
  title={Averaging methods for the asymptotic analysis of learning and adaptive systems, with small adjustment rate},
  author={Kushner, Harold Joseph and Huang, Hai},
  journal={SIAM Journal on Control and Optimization},
  volume={19},
  number={5},
  pages={635--650},
  year={1981},
  publisher={SIAM}
}

@article{hairer2010convergence,
  title={Convergence of Markov processes},
  author={Hairer, Martin},
  journal={Lecture notes},
  volume={18},
  number={26},
  pages={11},
  year={2010}
}

@article{Kryloff1937,
 ISSN = {0003486X, 19398980},
 URL = {http://www.jstor.org/stable/1968511},
 author = {Nicolas Kryloff and Nicolas Bogoliouboff},
 journal = {Annals of Mathematics},
 number = {1},
 pages = {65--113},
 publisher = {[Annals of Mathematics, Trustees of Princeton University on Behalf of the Annals of Mathematics, Mathematics Department, Princeton University]},
 title = {La Th\'{e}orie G\'{e}n\'{e}rale De La Mesure Dans Son Application \`{A} L'\'{E}tude Des Syst\'{e}mes Dynamiques De la M\'{e}canique Non Lin\'{e}aire},
 urldate = {2026-01-21},
 volume = {38},
 year = {1937}
}

@article{xu2024mechanism,
  title={On mechanism underlying algorithmic collusion},
  author={Xu, Zhang and Zhao, Wei},
  journal={arXiv preprint arXiv:2409.01147},
  year={2024}
}

@inproceedings{bertrand2025self,
  title={Self-Play {Q}-Learners Can Provably Collude in the Iterated Prisoner's Dilemma},
  author={Bertrand, Quentin and Duque, Juan Agustin and Calvano, Emilio and Gidel, Gauthier},
  booktitle={International Conference on Machine Learning},
  year={2025}
}

@article{banchio2022artificial,
  title={Artificial intelligence and spontaneous collusion},
  author={Banchio, Martino and Mantegazza, Giacomo},
  journal={arXiv preprint arXiv:2202.05946},
  year={2022}
}

@inproceedings{banchio2023adaptive,
author = {Banchio, Martino and Mantegazza, Giacomo},
title = {Adaptive Algorithms and Collusion via Coupling},
year = {2023},
isbn = {9798400701047},
publisher = {Association for Computing Machinery},
address = {New York, NY, USA},
url = {https://doi.org/10.1145/3580507.3597726},
doi = {10.1145/3580507.3597726},
booktitle = {Proceedings of the 24th ACM Conference on Economics and Computation},
pages = {208},
numpages = {1},
location = {London, United Kingdom},
series = {EC '23}
}

@article{carissimo2025algorithmic,
  title={Algorithmic Collusion is Algorithm Orchestration},
  author={Carissimo, Cesare and Falniowski, Fryderyk and Rahimi, Siavash and Nax, Heinrich},
  journal={arXiv preprint arXiv:2508.14766},
  year={2025}
}

@article{bladon2010evolutionary,
  title={Evolutionary dynamics, intrinsic noise, and cycles of cooperation},
  author={Bladon, Alex J and Galla, Tobias and McKane, Alan J},
  journal={Physical Review E},
  volume={81},
  number={6},
  pages={066122},
  year={2010},
  publisher={APS}
}

@article{barfuss2023intrinsic,
  title={Intrinsic fluctuations of reinforcement learning promote cooperation},
  author={Barfuss, Wolfram and Meylahn, Janusz M.},
  journal={Scientific Reports},
  volume={13},
  number={1},
  pages={1309},
  url={https://doi.org/10.1038/s41598-023-27672-7},
  doi={10.1038/s41598-023-27672-7},
  year={2023},
  publisher={Nature Publishing Group UK London}
}

@article{denBoer2022artificial,
  title = {Artificial Collusion: Examining Supracompetitive Pricing by {Q}-Learning Algorithms},
  author = {Arnoud V. den Boer and Janusz M. Meylahn and Maarten Pieter Schinkel},
  journal={Available at SSRN 4213600},
  doi={10.2139/ssrn.4213600},
  year={2024}
}

@article{meylahn2025quantifying,
  title={Quantifying the likelihood of learning collusive strategy equilibria},
  author={Meylahn, Janusz M},
  journal={Chaos: An Interdisciplinary Journal of Nonlinear Science},
  volume={35},
  number={8},
  year={2025},
  publisher={AIP Publishing}
}

@article{galla2011cycles,
  title={Cycles of cooperation and defection in imperfect learning},
  author={Galla, Tobias},
  journal={Journal of Statistical Mechanics: Theory and Experiment},
  volume={2011},
  number={08},
  pages={P08007},
  year={2011},
  doi={10.1088/1742-5468/2011/08/P08007},
  url={https://doi.org/10.1088/1742-5468/2011/08/P08007},
  publisher={IOP Publishing}
}

@article{meylahn2023does,
  title={Does an intermediate price facilitate algorithmic collusion?},
  author={Meylahn, Janusz M.},
  url={https://dx.doi.org/10.2139/ssrn.4594415},
  journal={Available at SSRN: 4594415},
  year={2023}
}

@article{meylahn2022limiting,
  title={Limiting dynamics for {Q}-learning with memory one in symmetric two-player, two-action games},
  author={Meylahn, Janusz M. and Janssen, Lars},
  journal={Complexity},
  volume={2022},
  pages={1--20},
  year={2022},
  url={https://doi.org/10.1155/2022/4830491},
  doi = {10.1155/2022/4830491},
  publisher={Hindawi Limited}
}

@article{abada2024algorithmic,
  title={Algorithmic Collusion: Where Are We and Where Should We Be Going?},
  author={Abada, Ibrahim and Harrington Jr, Joseph E. and Lambin, Xavier and Meylahn, Janusz M.},
  journal={Available at SSRN 4891033},
  year={2024}
}

@article{dolgopolov2024reinforcement,
  title={Reinforcement learning in a prisoner's dilemma},
  author={Dolgopolov, Arthur},
  journal={Games and Economic Behavior},
  volume={144},
  pages={84--103},
  year={2024},
  doi={10.1016/j.geb.2024.01.004},
  url={https://doi.org/10.1016/j.geb.2024.01.004},
  publisher={Elsevier}
}

@article{abada2024collusion,
  title={Collusion by mistake: Does algorithmic sophistication drive supra-competitive profits?},
  author={Abada, Ibrahim and Lambin, Xavier and Tchakarov, Nikolay},
  journal={European Journal of Operational Research},
  year={2024},
  doi={https://doi.org/10.1016/j.ejor.2024.06.006},
  url={https://doi.org/10.1016/j.ejor.2024.06.006},
  publisher={Elsevier}
}

@article{lambin2024less,
  title={Less than meets the eye: simultaneous experiments as a source of algorithmic seeming collusion},
  author={Lambin, Xavier},
  journal={Available at SSRN 4498926},
  year={2024}
}

@article{usui2021symmetric,
  title={Symmetric equilibrium of multi-agent reinforcement learning in repeated prisoner's dilemma},
  author={Usui, Yuki and Ueda, Masahiko},
  journal={Applied Mathematics and Computation},
  volume={409},
  pages={126370},
  year={2021},
  doi={10.1016/j.amc.2021.126370},
  publisher={Elsevier}
}

@article{epivent2024algorithmic,
  title={On algorithmic collusion and reward--punishment schemes},
  author={Epivent, Andr{\'e}a and Lambin, Xavier},
  journal={Economics Letters},
  volume={237},
  pages={111661},
  year={2024},
  publisher={Elsevier}
}

@article{klein2021autonomous,
  title={Autonomous algorithmic collusion: Q-learning under sequential pricing},
  author={Klein, Timo},
  journal={The RAND Journal of Economics},
  volume={52},
  number={3},
  pages={538--558},
  year={2021},
  publisher={Wiley Online Library}
}

@inproceedings{asker2022artificial,
  title={Artificial intelligence, algorithm design, and pricing},
  author={Asker, John and Fershtman, Chaim and Pakes, Ariel},
  booktitle={AEA Papers and Proceedings},
  volume={112},
  pages={452--456},
  year={2022},
  organization={American Economic Association 2014 Broadway, Suite 305, Nashville, TN 37203}
}

@article{asker2024impact,
  title={The impact of artificial intelligence design on pricing},
  author={Asker, John and Fershtman, Chaim and Pakes, Ariel},
  journal={Journal of Economics \& Management Strategy},
  volume={33},
  number={2},
  pages={276--304},
  year={2024},
  publisher={Wiley Online Library}
}

@article{calvano2020artificial,
  title={Artificial intelligence, algorithmic pricing, and collusion},
  author={Calvano, Emilio and Calzolari, Giacomo and Denicolo, Vincenzo and Pastorello, Sergio},
  journal={American Economic Review},
  volume={110},
  number={10},
  pages={3267--3297},
  year={2020},
  publisher={American Economic Association 2014 Broadway, Suite 305, Nashville, TN 37203}
}

@book{meyn2012markov,
  title={Markov chains and stochastic stability},
  author={Meyn, Sean P and Tweedie, Richard L},
  year={2012},
  publisher={Springer Science \& Business Media}
}

@article{devraj2017zap,
  title={Zap {Q}-learning},
  author={Devraj, Adithya M and Meyn, Sean},
  journal={Advances in Neural Information Processing Systems},
  volume={30},
  year={2017}
}

@book{Freidlin2012,
author="Freidlin, Mark I. and Wentzell, Alexander D.",
title="Random Perturbations of Dynamical Systems",
year="2012",
publisher="Springer Berlin Heidelberg",
address="Berlin, Heidelberg",
isbn="978-3-642-25847-3",
doi="10.1007/978-3-642-25847-3_1",
url="https://doi.org/10.1007/978-3-642-25847-3_1"
}

@article{waltman2008q,
  title={Q-learning agents in a Cournot oligopoly model},
  author={Waltman, Ludo and Kaymak, Uzay},
  journal={Journal of Economic Dynamics and Control},
  volume={32},
  number={10},
  pages={3275--3293},
  year={2008},
  publisher={Elsevier}
}

@article{calvano2021algorithmic,
  title={Algorithmic collusion with imperfect monitoring},
  author={Calvano, Emilio and Calzolari, Giacomo and Denicol{\'o}, Vincenzo and Pastorello, Sergio},
  journal={International journal of industrial organization},
  volume={79},
  pages={102712},
  year={2021},
  publisher={Elsevier}
}

@article{den2024mathematical,
  title={A (mathematical) definition of algorithmic collusion},
  author={den Boer, Arnoud V and Meylahn, Janusz M},
  journal={Available at SSRN 5012923},
  year={2024}
}

@article{assad2024algorithmic,
  title={Algorithmic pricing and competition: Empirical evidence from the German retail gasoline market},
  author={Assad, Stephanie and Clark, Robert and Ershov, Daniel and Xu, Lei},
  journal={Journal of Political Economy},
  volume={132},
  number={3},
  pages={723--771},
  year={2024},
  publisher={The University of Chicago Press Chicago, IL}
}

@article{brown2023competition,
  title={Competition in pricing algorithms},
  author={Brown, Zach Y and MacKay, Alexander},
  journal={American Economic Journal: Microeconomics},
  volume={15},
  number={2},
  pages={109--156},
  year={2023},
  publisher={American Economic Association 2014 Broadway, Suite 305, Nashville, TN 37203-2425}
}

@article{mackay2022dynamic,
  title={Dynamic pricing algorithms, consumer harm, and regulatory response},
  author={MacKay, Alexander and Weinstein, Samuel N},
  journal={Wash. UL Rev.},
  volume={100},
  pages={111},
  year={2022},
  publisher={HeinOnline}
}

@article{harrington2018developing,
  title={Developing competition law for collusion by autonomous artificial agents},
  author={Harrington, Joseph E},
  journal={Journal of Competition Law \& Economics},
  volume={14},
  number={3},
  pages={331--363},
  year={2018},
  publisher={Oxford University Press}
}

@inproceedings{hartline2024regulation,
  title={Regulation of algorithmic collusion},
  author={Hartline, Jason D and Long, Sheng and Zhang, Chenhao},
  booktitle={Proceedings of the 2024 Symposium on Computer Science and Law},
  pages={98--108},
  year={2024}
}

@article{abada2023artificial,
  title={Artificial intelligence: Can seemingly collusive outcomes be avoided?},
  author={Abada, Ibrahim and Lambin, Xavier},
  journal={Management Science},
  volume={69},
  number={9},
  pages={5042--5065},
  year={2023},
  publisher={INFORMS}
}
